\documentclass[english,aps,pra,twocolumn,showpacs]{revtex4-1}
\usepackage[T1]{fontenc}
\usepackage[latin9]{inputenc}
\usepackage{xcolor}
\usepackage{amsmath}
\usepackage{amssymb}
\usepackage{graphicx}
\usepackage{esint}
\PassOptionsToPackage{normalem}{ulem}
\usepackage{ulem}
\usepackage{babel}

\begin{document}
\title{Exact Non-Markovian Master Equation and Dispersive Probing of Non-Markovian
Process}

\author{Li-Ping Yang,$^{1}$ C. Y. Cai,$^{1}$ D. Z. Xu,$^{1}$
\\Wei-Min Zhang,$^{2}$ and C. P. Sun$^{1,3}$}
\selectlanguage{english}%

\affiliation{$^{1}$State Key Laboratory of Theoretical Physics, Institute of
Theoretical Physics, Chinese Academy of Science, Beijing 100190, China\\$^{2}$Department
of Physics and Center for Quantum Information Science, National Cheng
Kung University, Tainan 70101, Taiwan\\$^{3}$Beijing Computational
Science Research Center, Beijing 100084, China}
\begin{abstract}
For a bosonic (fermionic) open system in a bath with many bosons
(fermions) modes, we derive the exact non-Markovian master equation
in which the memory effect of the bath is reflected in
the time dependent decay rates. In this approach, the reduced
density operator is constructed from the formal
solution of the corresponding Heisenberg equations. As an application
of the exact master equation, we study the active probing of non-Markovianity
of the quantum dissipation of a single boson mode of electromagnetic
(EM) field in a cavity QED system. The non-Markovianity
of the bath of the cavity is explicitly reflected by the atomic decoherence
factor.
\end{abstract}


\maketitle
\section{INTRODUCTION}

The open quantum system approach is of much significance due to its
various applications in physics, e.g., quantum information, quantum transport,
and quantum chemistry, etc. Since a realistic quantum system is inevitably
coupled to many degrees of freedom in its environment
that leads to decoherence of the systems, a general approach to the
open quantum system is needed for its dissipative and dephasing process.
The dynamics of an open system is conventionally described with three
approaches: effective Hamiltonian~\cite{eff_Hamiltonian1,eff_Hamiltonian2,CPSun1995,CPSun1998,LHYu1994},
quantum master equations~\cite{Breuer,master_equaiton}, and quantum
Langevin equations~\cite{Langevin_equation,Louisell}. The last two approaches
are both based on the modeling with system plus bath, while the first
one is phenomenologically given by a time-dependent or non-Hermitian
Hamiltonian, which could lead to the dissipative motion equations.

About twenty years ago, Yu and one (C. P. S.) of the authors revealed
an intrinsic relation between the effective Hamiltonian and quantum
Langevin equation, obtained from the Heisenberg equations~\cite{LHYu1994,CPSun1995}.
By discarding the quantum fluctuation for the wide wave packet, they
derived the effective Hamiltonian of the system through the formally
exact solution for the time-dependent wave function of the total system.
However, the resulting effective Hamiltonian ignores the memory
effect, which is induced by the back action of the bath with time
delay. Therefore, if one wanted to recover the non-Markovian phenomenon
with memory effect, the quantum fluctuation of the bath must be taken
into account in the above Heisenberg equation based approach. To this
end, we need start from the Heisenberg equations of the total system, which
can reflect the original role of the bath. In this paper, without
any approximation, we derive the exact non-Markovian master equation
of the system from the formal solution of the Heisenberg equations.
The non-Markovian effect is contained in the time-dependent decay
rates in a straightforward way~\cite{PRB08_WMZ}.

It is commonly believed that the Markov process happens when the system-bath
coupling is weak. However, with the rapid development of experimental
technology, the strong-coupling limit can be reached. The theory of
open quantum systems in the strong-coupling regime is required for a proper
description of the non-Markovian dynamics. Recently,
many works on exact quantum master equations have been done
~\cite{PRD1992,PRA04_TYu,PRA07_WMZ,AnnP12_WMZ,PRB08_WMZ,NJP10_JSJin,PRA10_WMZ,PRA10_Breuer,HFL}.
In particular, one (W. M. Z.) of the authors and his collaborators
derived the exact non-Markovian master equations with a Lindblad-form for both Bose~\cite{PRA07_WMZ,AnnP12_WMZ}
and Fermi~\cite{PRB08_WMZ,NJP10_JSJin} systems by a path-integral
method in coherent-state representation. We now revisit these non-Markovian
master equations by generalizing our previous approach~\cite{CPSun1998},
which was used to derive a partially factorized wave function for
open quantum system. Using the present generalization to derive
the reduced density matrix is quite straightforward. Here, we
first construct the total density matrix with the help of the formal
solution of the Heisenberg equations, and then trace over the degrees
of freedom of the bath to obtain the reduced density matrix of the
system in the coherent state representation, instead of using
the Feynman-Vernon influence functional, as was done in
Refs.~\cite{PRB08_WMZ,NJP10_JSJin,AnnP12_WMZ}.
It reproduces the same reduced density matrix that satisfies a
time-local master equation where the non-Markovian memory effect
is fully taken into account.

With the help of the exact reduced density matrix, the dynamics of
an open quantum system could be well described. Meanwhile, there are
several proposals to measure the degree of the non-Markovianity of
open quantum process~\cite{PRL09_Breuer,PRA10_Sun}.
Very recently, the general non-Markovian dynamics of the
environment on its surrounding open quantum system
are explored within the exact master equation~\cite{arXiv12_WMZhang}.
The question is how to probe the general non-Markovian dynamics.
We thereby propose in this
paper a promising approach to probe the time-dependent memory effect of a bath on a
damped micro-cavity by coupling the cavity to a two-level atom dispersively.
To probe the non-Markovianity of the dissipation of the single model
EM field in a cavity, we let atoms of large detuning pass through
the cavity. We found that the non-Markovianity of the bath
is explicitly reflected by the atomic decoherence factor.
In the week coupling region, the periodically reviving amplitude decreases
along with the cavity-bath coupling strength and decays to $0$ finally.
On the the contrary, in the strong coupling region, the reviving amplitude
increases with the coupling strength and almost does not decay in the
ultra-strong coupling case, as a significant non-Markovian effect~\cite{arXiv12_WMZhang}.
This atomic decoherence factor could be detected through the Ramsey
interference in experiments.

In the next section, we solve the Heisenberg equations of the unified
quantum system plus bath model (Bose and Fermi) and obtain their formal
solutions. In Sec.~III, the derivation of the exact master equation
of Bose system is presented. The exact master equation of Fermi case
is addressed in Sec.~IV. In Sec. V, we propose to probe the non-Markovian
dynamics of a damped cavity with largely detuned two-level atoms.
Finally, the summery of our main results is given in Sec. VI. Some
detailed calculations are displayed in the Appendices.

\section{UNIFIED QUANTUM BATH MODEL AND FORMAL SOLUTION OF THE HEISENBERG EQUATIONS}

We consider an open quantum system $S$, which interacts with another
large system $B$ called bath. The combined system $S+B$ is usually
assumed to be closed, thus regarded as a Universe. The coupling of
$S$ to $B$ will lead to the dissipation and dephasing of $S$. There
are various types of bath, but the most commonly employed baths
are modeled with non-interacting bosons and fermions. In this paper we consider the specific
cases: a Bose system is surrounded by a Bose bath, or a Fermi system
is immersed in a Fermi bath. Here, we first solve the Heisenberg equations
for both the Bose and Fermi cases and obtain their formally exact solutions.

The Universe Hamiltonian $H=H_{s}+H_{b}+H_{\mathrm{int}}$ is decomposed
into three parts: the Hamiltonian of the system is taken to be a quadratic
form
\begin{equation}
H_{s}=\left[a_{1}^{\dagger},a_{2}^{\dagger},\cdots,a_{N_{s}}^{\dagger}\right]M\left[a_{1},a_{2},\dots,a_{N_{s}}\right]^{T},
\end{equation}
which describes $N_{s}$ linearly coupled bosons or fermions. $a_{i}(a_{i}^{\dagger})$
is the annihilation (creation) operator of the $i$th mode of the
system satisfying the commutation relation $\left[a_{i},a_{i'}^{\dagger}\right]_{\mp}=\delta_{ii'}$
($\mp$ corresponds to the boson and fermion, respectively) and $M$
is a positive definite Hermitian matrix. The Hamiltonian of the Bose
or Fermi bath is given by
\begin{equation}
H_{b}=\sum_{l=1}^{N_{b}}\omega_{l}b_{l}^{\dagger}b_{l},
\end{equation}
with the number of the uncoupled modes of the bath $N_{b}(\gg N_{s})$
and annihilation (creation) operators $b_{l}$ ($b_{l}^{\dagger}$)
which satisfy corresponding commutation relations $\left[b_{l},b_{l'}^{\dagger}\right]_{\mp}=$$\delta_{ll'}$.
As proofed in~\cite{master_equaiton}, the most usual environment coupled to the open system could be well approximated as a collection of harmonic oscillators with linear quadratic couplings. Here, the interaction Hamiltonian is taken as the
form of
\begin{equation}
H_{\mathrm{int}}=\sum_{i=1}^{N_{s}}\sum_{l=1}^{N_{b}}\left(\eta_{il}a_{i}^{\dagger}b_{l}+\eta_{il}^{*}b_{l}^{\dagger}a_{i}\right).
\end{equation}

In the Heisenberg picture, the dynamics of the system is governed
by the Heisenberg equations:
\begin{eqnarray}
\dot{a}_{i}\left(t\right) & = & -\mathrm{i}\sum_{j}M_{ij}a_{j}\left(t\right)-\mathrm{i}\sum_{l}\eta_{il}b_{l}\left(t\right),\label{eq:HE1}\\
\dot{b}_{l}\left(t\right) & = & -\mathrm{i}\omega_{l}b_{l}\left(t\right)-\mathrm{i}\sum_{i}\eta_{li}^{*}a_{i}\left(t\right).\label{eq:HE2}
\end{eqnarray}
 For convenience, we introduce the $\left(N_{s}+N_{b}\right)$ operator-valued
vector
\[
\vec{c}\left(t\right)\!\!=\!\!\left[\!\vec{a},\vec{b}\right]^{T}\!\!\!\!=\!\!\left[a_{1}\!\!\left(\! t\!\right)\!,a_{2}\!\!\left(\! t\!\right)\!,\!\cdots\!,a_{N_{s}}\!\!\left(\! t\!\right)\!,b_{1}\!\!\left(\! t\!\right)\!,b_{2}\!\!\left(\! t\!\right)\!,\!\cdots\!,b_{N_{b}}\!\!\left(\! t\!\right)\right]^{T}\!\!\!\!\!,
\]
and the $\left(N_{s}+N_{b}\right)\times\left(N_{s}+N_{b}\right)$
coefficient matrix
\begin{equation}
\mathcal{H}=\left[\begin{array}{cc}
M & R\\
R^{\dagger} & E
\end{array}\right],\label{eq:M_tilde}
\end{equation}
where
\[
R=\left[\begin{array}{cccc}
\eta_{11} & \eta_{12} & \dots & \eta_{1N_{b}}\\
\eta_{21} & \eta_{22} & \cdots & \eta_{2N_{b}}\\
\vdots & \vdots & \vdots & \vdots\\
\eta_{N_{s}1} & \eta_{N_{s}2} & \cdots & \eta_{N_{s}N_{b}}
\end{array}\right],
\]
and
\[
E=\mathrm{diag}\left[\omega_{1},\omega_{2},\cdots,\omega_{N_{b}}\right].
\]
Then Eqs.~(\ref{eq:HE1}) and (\ref{eq:HE2}) are re-expressed in
a compact form
\begin{equation}
\frac{d}{dt}\vec{c}\left(t\right)=-\mathrm{i}\mathcal{H}\vec{c}\left(t\right).\label{eq:HE-matrix_form}
\end{equation}
It follows from Eq.~(\ref{eq:M_tilde}) that $\mathcal{H}$ is a
time-independent Hermitian matrix. Consequently, the formal solution
of Eq.~(\ref{eq:HE-matrix_form}) is given by
\[
\vec{c}\left(t\right)=\exp\left[-\mathrm{i}\mathcal{H}t\right]\vec{c}\left(0\right)\equiv\mathcal{U}\left(t\right)\vec{c}\left(0\right),
\]
where $\mathcal{U}\left(t\right)=\exp\left[-\mathrm{i}\mathcal{H}t\right]$
is the time-evolution operator. Splitting the matrix $\mathcal{U}\left(t\right)$
into four blocks
\begin{equation}
\mathcal{U}\left(t\right)\equiv\left[\begin{array}{cc}
\left[W\left(t\right)\right]_{N_{s}\times N_{s}} & \left[T\left(t\right)\right]_{N_{s}\times N_{b}}\\
\left[P\left(t\right)\right]_{N_{b}\times N_{s}} & \left[Q\left(t\right)\right]_{N_{b}\times N_{b}}
\end{array}\right],\label{eq:evolution_operator}
\end{equation}
we obtain formal solution of Eq.~(\ref{eq:HE-matrix_form}) as
\begin{eqnarray}
\vec{a}\left(t\right) & = & W\left(t\right)\vec{a}\left(0\right)+T\left(t\right)\vec{b}\left(0\right),\label{eq:solu1}\\
\vec{b}\left(t\right) & = & P\left(t\right)\vec{a}\left(0\right)+Q\left(t\right)\vec{b}\left(0\right).\label{eq:solu2}
\end{eqnarray}

The dynamics of total system is governed by these four time-dependent
coefficient matrices $W\left(t\right)$, $T\left(t\right)$, $Q\left(t\right)$,
and $P\left(t\right)$.

Up to now, all the results are obtained by formal operations, since
these coefficient matrices need to be determined by the differential
equations. As shown in Appendix B, there are some connections between
these coefficient matrices, which take a crucial role in derivation
of the exact master equation.

\subsection{Differential equations of the coefficient matrices}

Substituting Eqs.~(\ref{eq:solu1}) and (\ref{eq:solu2}) into Eqs.~(\ref{eq:HE1})
and (\ref{eq:HE2}), we obtain the equations of the coefficient matrices
\begin{eqnarray}
\dot{W}\left(t\right) & = & -\mathrm{i}\left[MW\left(t\right)+RP\left(t\right)\right],\label{eq:W_dot}\\
\dot{T}\left(t\right) & = & -\mathrm{i}\left[MT\left(t\right)+RQ\left(t\right)\right],\label{eq:T_dot}\\
\dot{P}\left(t\right) & = & -\mathrm{i}\left[EP\left(t\right)+R^{\dagger}W\left(t\right)\right],\\
\dot{Q}\left(t\right) & = & -i\left[EQ\left(t\right)+R^{\dagger}T\left(t\right)\right],
\end{eqnarray}
with the initial conditions
\begin{equation}
W\left(0\right)=I,\ T\left(0\right)=\mathbf{0},\ P\left(0\right)=\mathbf{0},\ Q\left(0\right)=I.
\end{equation}
Here, $I$ is the identity matrix and $\mathbf{0}$ is the null matrix.
The differential equations of $P\left(t\right)$ and $Q\left(t\right)$
are integrated to yield
\begin{eqnarray}
P\left(t\right) & = & -\mathrm{i}\int_{0}^{t}d\tau e^{-\mathrm{i}E\left(t-\tau\right)}R^{\dagger}W\left(\tau\right)d\tau,\label{eq:P_W}\\
Q\left(t\right) & = & e^{-\mathrm{i}Et}\left[-\mathrm{i}\int_{0}^{t}d\tau e^{\mathrm{i}E\tau}R^{\dagger}T\left(\tau\right)+I\right].
\end{eqnarray}
Then, we obtain the integrodifferential equations about $W\left(t\right)$
and $T\left(t\right)$:
\begin{eqnarray}
\dot{W}\left(t\right)+\mathrm{i}MW\left(t\right)+\int_{0}^{t}d\tau G\left(t-\tau\right)W\left(\tau\right) & = & 0,\label{eq:W_integrodiff}
\end{eqnarray}
\begin{equation}
\dot{T}\left(t\right)+\mathrm{i}MT\left(t\right)+\int_{0}^{t}d\tau G\left(t-\tau\right)T\left(\tau\right)=-\mathrm{i}Re^{-\mathrm{i}Et}\!.\label{eq:T_integrodiff}
\end{equation}
Here the ($N_{s}\times N_{s}$) kernel matrix $G\left(t\right)=Re^{-\mathrm{i}Et}R^{\dagger}$
characterizes the non-Markovian memory structure of $S$. Defining the
interaction spectral function
\[
J_{ij}\left(\omega\right)=\sum_{l}\eta_{il}\eta_{lj}^{*}\delta\left(\omega-\omega_{l}\right),
\]
we rewrite the element of the kernel matrix $G\left(t\right)$ as
\[
G_{ij}\left(t\right)=\int d\omega J_{ij}\left(\omega\right)e^{-i\omega t}.
\]
Thus, the matrix $G\left(t\right)$ is
fully determined by the interaction spectrum.

On the other hand, the coefficient matrices $W\left(t\right)$ and
$T\left(t\right)$ are not independent. By taking the Laplace transform
of the integral differential equations (\ref{eq:W_integrodiff}) and
(\ref{eq:T_integrodiff}), we get
\begin{eqnarray}
W[p] & = & \mathcal{L}\left(W\right)=\left[p+\mathrm{i}M+G\left(p\right)\right]^{-1},\\
T[p] & = & W[p]\mathcal{L}\left(-\mathrm{i}Re^{-\mathrm{i}Et}\right),
\end{eqnarray}
where $\mathcal{L}\left(\cdots\right)$ represents the Laplace transform.
Consequently, after the inverse Laplace transform, the matrix $T\left(t\right)$
is given by
\begin{equation}
T\left(t\right)=-\mathrm{i}\int_{0}^{t}d\tau W\left(t-\tau\right)Re^{-\mathrm{i}E\tau}.\label{eq:T_W}
\end{equation}

Thus the dynamics of $S$ could be completely described by a single
coefficient matrix $W\left(t\right)$,
It is well known that, under
the Wigner-Weisskopff approximation, one can obtain the quantum Langevin
equations of the operators of $S$ by means of the approximate solution
of Eqs.(\ref{eq:W_integrodiff}) and (\ref{eq:T_integrodiff}) together
with the Heisenberg equations (\ref{eq:HE1}) and (\ref{eq:HE2})~\cite{Louisell}.
In this paper, it will be shown that the exact master equation of
the reduced density matrix can also be obtained based on the formal
solutions (\ref{eq:solu1}-\ref{eq:solu2}) of the Heisenberg equations.
And the Wigner-Weisskopff approximation leads to the quantum Born-Markov
master equation.

\section{BOSON CASE IN COHERENT-STATE REPRESENTATION}

In this section, we derive the exact master equation for $N_{s}$
coupled bosons in a Bose bath. In the Schr$\ddot{\mathrm{o}}$dinger
picture, the total density matrix $\rho\left(t\right)=U\left(t\right)\rho\left(0\right)U^{\dagger}\left(t\right)$
of $S+B$ obeys the Liouville-von Neumann equation $\mathrm{i}\hbar\dot{\rho}\left(t\right)=\left[H,\rho\left(t\right)\right]$,
where $U\left(t\right)=\exp\left(-\mathrm{i}Ht\right)$ is the time
evolution operator of the total system. We assume that the total system
is initially in the direct product initial state $\rho\left(0\right)=\rho_{s}\left(0\right)\otimes\rho_{b}\left(0\right)$,
with density matrices $\rho_{s}\left(0\right)$ and $\rho_{b}\left(0\right)$
of $S$ and $B$, respectively. Through a lengthy
calculation in Appendix C, the reduced density matrix of $S$ is expressed
in terms of the coherent state $\left|\vec{x}\right\rangle $ of the
system
\begin{eqnarray}
\rho_{s}\left(t\right) & = & \int d\mu\left(\vec{\alpha},\vec{\alpha}^{\prime}\right)d\mu\left(\vec{\xi},\vec{\xi}^{\prime}\right)\left|\vec{\alpha}\right\rangle \!\left\langle \vec{\alpha}^{\prime}\right|\nonumber \\
 &  & K\left(\vec{\alpha}^{\dagger},\vec{\alpha}^{\prime},\vec{\xi},\vec{\xi}^{\prime\dagger},t\right)\left\langle \vec{\xi}\right|\rho_{s}\left(0\right)\left|\vec{\xi}^{\prime}\right\rangle ,\label{eq:reduced_density_m}
\end{eqnarray}
with $\vec{x}=\left[x_{1},x_{2},\cdots,x_{N_{s}}\right]^{T}$ ($\vec{x}=\vec{\alpha},\vec{\alpha}^{\prime},\vec{\xi},\vec{\xi}^{\prime}$).
The propagator, which governs the dynamics of the reduced density
matrix, is defined as
\begin{eqnarray}
K\left(\vec{\alpha}^{\dagger},\vec{\alpha}^{\prime},\vec{\xi},\vec{\xi}^{\prime\dagger};t\right) & = & \int d\mu\left(\vec{z}\right)\left\langle \vec{\alpha},\vec{z}\right|U\left(t\right)\left|\vec{\xi}\right\rangle \nonumber \\
 &  & \!\times\!\left\langle \vec{\xi}^{\prime}\right|\rho_{b}\left(0\right)U^{\dagger}\left(t\right)\left|\vec{\alpha}^{\prime},\vec{z}\right\rangle .\label{eq:propagating_f}
\end{eqnarray}
Here $\left|\vec{z}\right\rangle $$\left(\vec{z}=\left[z_{1},z_{2},\cdots,z_{N_{b}}\right]\right)$
is the coherent state of $B$.
Different from the previous derivation~\cite{PRB08_WMZ,NJP10_JSJin,AnnP12_WMZ} where the
propagating function is obtained using the coherent-state path integral method and tracing
over the environmental degrees of freedom completely through the Feynman-Vernon influence functional,
the propagator could also be evaluated in the coherent state
representation by constructing the explicit total wave
function~\cite{CPSun1998}
\begin{equation}
U^{\dagger}\left(t\right)\left|\vec{\alpha}^{\prime},\vec{z}\right\rangle =\exp\left[\vec{a}^{\dagger}\left(t\right)\cdot\vec{\alpha}^{\prime}+\vec{b}^{\dagger}\left(t\right)\cdot\vec{z}\right]\left|0\right\rangle , \label{eq:wave_function}
\end{equation}
as shown in Appendix C.
It deserves to be noted that we have used the identities $U^{\dagger}\left(t\right)\left|0\right\rangle =\left|0\right\rangle $
and $O\left(t\right)=U^{\dagger}\left(t\right)OU\left(t\right)$.

\subsection{Propagating Function}

Generally speaking, the bath is initially in its thermal equilibrium
state
\begin{equation}
\rho_{b}\left(0\right)=\left(\prod_{l}\frac{1}{f_{l}+1}\right)\exp\left[-\beta\vec{b}^{\dagger}E\vec{b}\right],\label{eq:ro_b}
\end{equation}
where $f_{l}=1/\left[\exp\left(\beta\omega_{l}\right)-1\right]$ is
the mean occupation number of the $l$th bath mode at temperature
$T=1/(k_{B}\beta)$. In this case, the integral over the bath in the
propagator (\ref{eq:propagating_f}) is carried out to give (please
refer to Appendix C for the detail),
\begin{eqnarray}
\!\!\!\!\!\!\!\!\!\!\!\!\!K(\vec{\alpha}^{\dagger},\vec{\alpha}^{\prime},\vec{\xi},\vec{\xi}^{\prime\dagger},t)\! & = & \!A(t)\!\exp\!\!\left[\!\!\begin{array}{c}
\vec{\alpha}^{\dagger}J_{1}(t)\vec{\xi}+\vec{\xi}^{\prime\dagger}J_{1}^{\dagger}(t)\vec{\alpha}^{\prime}\\
+\vec{\alpha}^{\dagger}J_{2}(t)\vec{\alpha}^{\prime}+\vec{\xi}^{\prime\dagger}J_{3}\vec{\xi}\!\!\!\!
\end{array}\right]\!,\label{eq:K}
\end{eqnarray}
where
\[
A\left(t\right) = \mathrm{det}\left[[I+V\left(t\right)]^{-1}\right],
 \]
\[
J_{1}\left(t\right) = [I+V\left(t\right)]^{-1}W\left(t\right),
\]
\[
J_{2}\left(t\right) = V[I+V\left(t\right)]^{-1},
\]
\[
J_{3}\left(t\right) =  I-W^{\dagger}\left(t\right)[I+V\left(t\right)]^{-1}W\left(t\right).
\]
This reproduces the propagating function obtained by the coherent state path-integral
method in the previous works, e.g., Eq.~(31) in~\cite{AnnP12_WMZ}. For convenience, we have introduced a new $N_{s}\times N_{s}$ Hermitian
matrix $V\left(t\right)=T\left(t\right)fT^{\dagger}\left(t\right)$.
Utilizing the relationship in Eq.~(\ref{eq:T_W}) between the matrices
$T\left(t\right)$ and $W\left(t\right)$, we have
\begin{equation}
V\left(t\right)=\int_{0}^{t}d\tau_{1}\int_{0}^{t}d\tau_{2}W\left(\tau_{1}\right)\tilde{G}\left(\tau_{2}-\tau_{1}\right)W^{\dagger}\left(\tau_{2}\right),
\end{equation}
with
\begin{equation}
\tilde{G}\left(t\right)=Rfe^{-\mathrm{i}Et}R^{\dagger}.
\end{equation}

Without any additional hypothesis, the exact propagating function
of the reduced density matrix of $S$ is obtained. The dynamics of
$S$ is governed by the single coefficient matrix $W\left(t\right)$,
which is determined by integral differential equation (\ref{eq:W_integrodiff}).
And the influence of the bath on the dynamics of $S$ is characterized
by two memory-kernel matrices $G\left(t\right)$ and
$\tilde{G}\left(t\right)$~\cite{PRB08_WMZ,NJP10_JSJin,AnnP12_WMZ}.
\vspace{0.6cm}
\subsection{The Exact Non-Markovian Master equation for Bosons}

In the proceeding subsection, we have obtained the exact reduced density
matrix of $S$ as in Eq.~(\ref{eq:reduced_density_m}). Now we construct
the master equation through its time derivative
\begin{eqnarray}
\dot{\rho}_{s} & = & \int d\mu\left(\vec{\alpha},\vec{\alpha}^{\prime}\right) d\mu\left(\vec{\xi},\!\vec{\xi}^{\prime}\right)\left|\vec{\alpha}\right\rangle \left\langle \vec{\alpha}^{\prime}\right| \nonumber \\
 &  & \dot{K}\!\!\left(\!\vec{\alpha}^{\dagger},\vec{\alpha}^{\prime},\vec{\xi},\vec{\xi}^{\prime\dagger}; t\right)\left\langle \vec{\xi}\right|\rho_{s}\left(0\right)\left|\vec{\xi}^{\prime}\right\rangle.
\end{eqnarray}
And it is found that the time differential of the propagating function
takes the following form (please refer to Appendix D for the detail)
\begin{eqnarray}
\!\!\!\dot{K} & = & \vec{\alpha}^{\dagger}\tilde{\Gamma}K\vec{\alpha}^{\prime}-\mathrm{Tr}\left[\tilde{\Gamma}\right]K-\vec{\alpha}^{\dagger}\left(\Gamma+\mathrm{i}\tilde{\Omega}+\tilde{\Gamma}\right)\overset{\rightarrow}{\nabla}_{\alpha^{*}}K\nonumber \\
 &  & \!\!\!\!-\!\left(\!\overset{\rightarrow}{\nabla}_{\alpha'}^{T}\! K\right)\!\left(\Gamma\!-\!\mathrm{i}\tilde{\Omega}\!+\!\tilde{\Gamma}\right)\!\vec{\alpha}^{\prime}\!+\!\overset{\rightarrow}{\nabla}_{\alpha'}^{T}\!\left(\tilde{\Gamma}\!+\!2\Gamma\right)\!\!\overset{\rightarrow}{\nabla}_{\alpha^{*}}K,
\end{eqnarray}
with Hermitian matrices
\begin{equation}
\tilde{\Gamma}\left(t\right)=\dot{V}\left(t\right)-\dot{W}\left(t\right)W^{-1}\left(t\right)V\left(t\right)-V\left(\dot{W}\left(t\right)W^{-1}\left(t\right)\right)^{\dagger},\label{eq:Gamma_t}
\end{equation}
\begin{equation}
\Gamma\left(t\right)=-\frac{1}{2}\left[\dot{W}\left(t\right)W^{-1}\left(t\right)+\left(\dot{W}\left(t\right)W^{-1}\left(t\right)\right)^{\dagger}\right],\label{eq:Gamma}
\end{equation}
and
\begin{equation}
\tilde{\Omega}\left(t\right)=\frac{\mathrm{i}}{2}\left[\dot{W}\left(t\right)W^{-1}\left(t\right)-\left(\dot{W}\left(t\right)W^{-1}\left(t\right)\right)^{\dagger}\right].\label{eq:Omega_t}
\end{equation}
For coherent state defined in Eq.~(\ref{eq:coherent_state}), there
exist the following relations~\cite{PRA1994_correspondence}
\begin{eqnarray*}
\vec{\alpha}\left|\vec{\alpha}\right\rangle  & = & \vec{a}\left|\vec{\alpha}\right\rangle ,\ \vec{\alpha}^{\dagger}\left\langle \vec{\alpha}\right|=\left\langle \vec{\alpha}\right|\vec{a}^{\dagger},\\
\overset{\rightarrow}{\nabla}_{\alpha}^{T}\left|\alpha\right\rangle  & = & \vec{a}^{\dagger}\left|\vec{\alpha}\right\rangle ,\ \overset{\rightarrow}{\nabla}_{\alpha^{*}}\left\langle \vec{\alpha}\right|=\left\langle \vec{\alpha}\right|\vec{a}.
\end{eqnarray*}
With these mapping, we can construct the exact master equation of
the reduced density matrix of the Bose system $S$, i.e. Eq.~(32) in Ref.~\cite{AnnP12_WMZ},\begin{widetext}
\begin{eqnarray}
\dot{\rho}_{s}\left(t\right)\! & = & -\mathrm{i}\left[\tilde{H}_{s}\left(t\right),\rho_{s}\left(t\right)\right]+\sum_{ij}\left[\tilde{\Gamma}_{ij}\left(t\right)+2\Gamma_{ij}\left(t\right)\right]\left[a_{j}\rho_{s}\left(t\right)a_{i}^{\dagger}-\frac{1}{2}a_{i}^{\dagger}a_{j}\rho_{s}\left(t\right)-\frac{1}{2}\rho_{s}a_{i}^{\dagger}a_{j}\right]\nonumber \\
 &  & +\sum_{ij}\tilde{\Gamma}_{ij}\left(t\right)\left[a_{i}^{\dagger}\rho_{s}\left(t\right)a_{j}-\frac{1}{2}a_{j}a_{i}^{\dagger}\rho_{s}\left(t\right)-\frac{1}{2}\rho_{s}\left(t\right)a_{j}a_{i}^{\dagger}\right],
\end{eqnarray}
\end{widetext}where $\tilde{H}_{s}=\vec{a}^{\dagger}\tilde{\Omega}\vec{a}$
is the effective time-dependent Hamiltonian of the system $S$. The
diagonal elements of $\tilde{\Omega}\left(t\right)$ are the modified
time-dependent frequencies of the different modes of $S$ and the
off-diagonals represent the new interaction strength between the modes
of the system. Without Markov approximation, the dissipation of the
system and the fluctuation of the bath could not be separated. The
original role of the bath is reflected by the time-dependent decay
coefficients $\Gamma\left(t\right)$ and
$\tilde{\Gamma}\left(t\right)$~\cite{AnnP12_WMZ}.

\subsection{From Wigner-Weisskopff Approximation to Markov Master Equation}

In this subsection, it will be shown that the Markov master equation can be
obtained from the exact master equation by taking Wiger-Weisskopff
approximation~\cite{Louisell}, instead of making a direct Markov
approximation~\cite{PRA10_WMZ}. Here the exact
master equation is applied to the simplest dissipative system consisting
of a single harmonic oscillator with frequency $\Omega_{0}$ and a
Bose environment. In this case, $\tilde{\Omega}$, $\Gamma\left(t\right)$,
and $\tilde{\Gamma}\left(t\right)$ are just time-dependent numbers
instead of matrices, which are all determined by $W\left(t\right)$
in Eqs.~(\ref{eq:Gamma_t}-\ref{eq:Omega_t}). Under the Wigner-Weisskopff
approximation, the solution of Eq.~(\ref{eq:W_integrodiff}) is given
by
\begin{equation}
W(t)=\exp\left[-\Gamma_{0}t-i(\Omega_{0}+\Delta\omega)t\right],
\end{equation}
where
\begin{equation}
\Gamma_{0}=\pi J\left(\Omega_{0}\right),
\end{equation}
is the decay rate of the oscillator induced by the coupling to the
vacuum and
\begin{equation}
\Delta\omega=-\mathcal{P}\int\frac{J\left(\Omega_{0}\right)}{\omega-\Omega_{0}}d\omega,
\end{equation}
is the small frequency shift, with the interaction spectrum $J\left(\omega\right)$.
It is easy to find that, in this case, the parameters of
the master equation become time-independent
\begin{equation}
\tilde{\Omega}=\Omega_{0}+\Delta\omega,\ \Gamma=\Gamma_{0},\ \tilde{\Gamma}=2f\left(\Omega_{0}\right)\gamma_{0},
\end{equation}
where $f\left(\Omega_{0}\right)$ is the mean occupation number of
the oscillator. As we know, $\Gamma$ characterizes the dissipation
of the system and $\tilde{\Gamma}$ corresponds to the fluctuation
of the bath.

Then the Born-Markov master equation of a damped harmonic resonator
is obtained as\begin{widetext}
\begin{align}
\dot{\rho}_{s}(t)= & -i[\tilde{\Omega}a^{\dagger}a,\rho_{s}(t)]+\left[1+f(\Omega_{0})\right]\Gamma_{0}\left[2a\rho_{s}(t)a^{\dagger}-\left(a^{\dagger}a\rho_{s}(t)+\rho_{s}(t)a^{\dagger}a\right)\right]\nonumber \\
 & +f(\Omega_{0})\Gamma_{0}\left[2a^{\dagger}\rho_{s}(t)a-\left(aa^{\dagger}\rho_{s}(t)-\rho_{s}(t)aa^{\dagger}\right)\right].
\end{align}
\end{widetext}It is known that, for a damped harmonic oscillator, the
quantum Langevin equation of the number operator obtained from the
Markov approximation is same as the one from the Wigner-Weisskopff
approximation~\cite{Louisell}. In this sense, these two approximations
are equivalent.

\section{FERMI CASE IN COHERENT STATE REPRESENTATION}

In the previous section, we obtained the exact master equation of
the Bose system. Analogously, in the case of Fermi system, the reduced
density matrix in the fermion coherent state representation~\cite{fermion_coherent1,fermion_coherent2}
reads
\begin{eqnarray}
\rho_{s}\left(t\right) & = & \int d\mu\left(\vec{\alpha},\vec{\alpha}^{\prime}\right)d\mu\left(\vec{\xi},\vec{\xi}^{\prime}\right)\left|\vec{\alpha}\right\rangle \left\langle \vec{\alpha}^{\prime}\right|\nonumber \\
 &  & \times K\left(\vec{\alpha}^{\dagger},\vec{\alpha}^{\prime},\vec{\xi},\vec{\xi}^{\prime\dagger},t\right)\left\langle \vec{\xi}\right|\rho_{s}\left(0\right)\left|\vec{\xi}^{\prime}\right\rangle ,
\end{eqnarray}
where the components of vectors $\vec{\alpha}$, $\vec{\alpha}^{\prime}$,
$\vec{\xi}$, and $\vec{\xi}^{\prime}$ are Grassmann variables, $\rho_{s}\left(0\right)$
is the initial state of $S$. And the initial state of the bath is
still assumed to be the thermal state
\begin{equation}
\rho_{b}\left(0\right)=\prod_{l}\left(1-f_{l}\right)\exp\left[-\beta\vec{b}^{\dagger}E\vec{b}\right].
\end{equation}
where $f_{k}=1/\left[\exp\left(\beta\omega_{l}\right)+1\right]$ is
the mean occupation number of the $l$th Fermi mode with $\beta=1/\left(k_{B}T\right)$.
After tracing over the degrees of freedom of the bath, we find that
the propagator is of the same form as the Bose case~\cite{PRB08_WMZ},
\[
K=A\exp\left[\vec{\alpha}^{\dagger}J_{1}\vec{\xi}+\vec{\xi}^{\prime}J_{1}^{\dagger}\vec{\alpha}^{\prime}+\vec{\alpha}J_{2}\vec{\alpha}^{\prime}+\vec{\xi}^{\prime\dagger}J_{3}\vec{\xi}\right],
\]
but the matrices in $K$ change into
\[
A\left(t\right)=\mathrm{det}\left[[I-V\left(t\right)]^{-1}\right],
\]
\[
J_{1}\left(t\right)=\left[I-V\left(t\right)\right]^{-1}W\left(t\right),
\]
\[
J_{2}\left(t\right)=\left[I-V\left(t\right)\right]^{-1}-I,
\]
\[
J_{3}\left(t\right)=W^{\dagger}\left(t\right)\left[I-V\left(t\right)\right]^{-1}W\left(t\right)-I.
\]

After the same procedure as the Bose system, the exact master equation
of the Fermi system is obtained as the same one given by Eq.~(8) in~\cite{NJP10_JSJin} \begin{widetext}
\begin{eqnarray}
\dot{\rho}_{s}\left(t\right) & = & -\mathrm{i}\left[\tilde{H}_{s}\left(t\right),\rho_{s}\left(t\right)\right]+\sum_{ij}\left[2\Gamma_{ij}\left(t\right)-\tilde{\Gamma}_{ij}\left(t\right)\right]\left[a_{j}\rho_{s}\left(t\right)a_{i}^{\dagger}-\frac{1}{2}a_{i}^{\dagger}a_{j}\rho_{s}\left(t\right)-\frac{1}{2}\rho_{s}\left(t\right)a_{i}^{\dagger}a_{j}\right]\nonumber \\
 &  & +\sum_{ij}\tilde{\Gamma}_{ij}\left(t\right)\left[a_{i}^{\dagger}\rho_{s}\left(t\right)a_{j}-\frac{1}{2}a_{j}a_{i}^{\dagger}\rho_{s}\left(t\right)-\frac{1}{2}\rho_{s}\left(t\right)a_{j}a_{i}^{\dagger}\right],
\end{eqnarray}
\end{widetext}where $\tilde{H}_{s}\left(t\right)$, $\Gamma\left(t\right)$,
and $\tilde{\Gamma}\left(t\right)$ are defined in the same way as
the bosons'~\cite{PRB08_WMZ,NJP10_JSJin}.

\section{PROBING NON-MARKOVIANIANITY OF AN OPEN QUANTUM SYSTEM}

In this section, we consider how to probe the non-Markovianity of
a quantum dissipation process in a realistic physical system. We understand
that such an ideal probing scheme is usually based on the non-demolition
measurement\cite{Quantum_Measurement}. The interaction between the
probing apparatus and the system to be detected commutes with the
free Hamiltonian of the system, thus such kind of measurement does
not change the energy of the system. But it will retain the information
of the system on the probing apparatus. Such non-demolition interaction
can be implemented in the cavity-QED as the dispersive interaction
between the atom and cavity \cite{PRL90_QND,PRA92_QND}. On the other
hand, it is feasible to prepare and analyze a two-level Rydberg atom
in a state corresponding to an arbitrary point on the Bloch sphere
in the quantum optics experiments.

To realize the probing non-Markovianity in the cavity QED system,
we consider an open quantum system: a single cavity mode coupled to
its bath of many bosonic excitation modes resulting from the cavity
leakage. Let an atom pass through the cavity, and then examine
the quantum coherence of the atom. In this case, the atom could record
the intrinsic information of the cavity field to accomplish the probing
of the non-Markovianity of the cavity dynamics. This kind of approach
was also used to probe the quantum criticality of many body system
\cite{PRL06_HTQuan}, where the sensitive change of the atom decoherence
factor, which is characterized by the Loschmidt echo~\cite{Lochsmidt},
could reflect the quantum criticality of its surrounding environment.

\begin{figure}
\includegraphics[width=8cm]{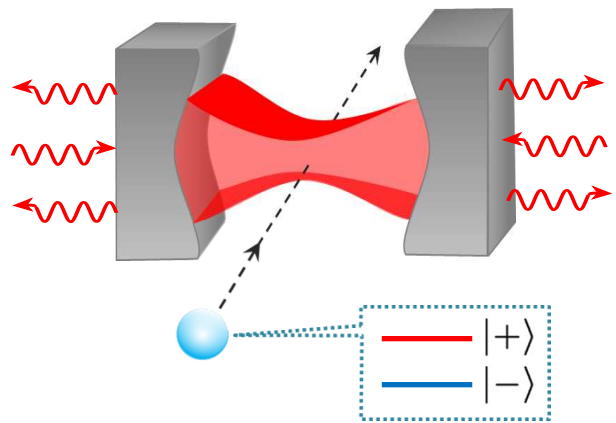}

\caption{(Color online) Schematic diagram for probing of the non-Markovian
dynamics of an open quantum system: a leaking cavity. The two-level
atom passing through the cavity is largely detuned from the frequency
of the cavity mode to approach the non-demolition measurement.}
\end{figure}

In our case, the frequency of the atom $\omega_{a}$ is drastically
detuned from the the cavity resonance frequency $\omega_{0}$, i.e.,
$\Delta=\omega_{0}-\omega_{a}\gg g_{a-f}$, where $g_{a-f}$ is the
vacuum Rabi frequency characterizing the atom-cavity coupling. By
making use of an adiabatic elimination procedure, we obtain the effective
Hamiltonian
\begin{equation}
H_{p}=\hbar\omega_{0}a^{\dagger}a+\hbar\omega_{a}\sigma_{z}+\hbar\delta a^{\dagger}a\sigma_{z},
\end{equation}
for our probing scheme from the usual Jaynes-Cummings model~\cite{Scully}.
Here, $a(a^{\dagger})$ is the annihilation (creation) operator of
the cavity, $\sigma_{z}=\left|e\right\rangle \left\langle e\right|-\left|g\right\rangle \left\langle g\right|$
is the Pauli matrix of the atom with the ground (excited) state of
the atom $\left|g\right\rangle (\left|e\right\rangle )$, and $\delta=g_{a-f}^{2}/\Delta$
is the effective dispersive coupling constant~\cite{PRL1996_Haroche,PRL1997_Haroche}.
Meanwhile, the cavity is coupled to a bosonic bath
\[
H_{b}+H_{\mathrm{int}}=\sum_{l}\hbar\omega_{l}b_{l}^{\dagger}b_{l}+\hbar\sum_{l}\left(\eta_{l}a^{\dagger}b_{l}+\mathrm{H.c.}\right).
\]
Here, the atom has enough long coherence time and we neglect the decay
of the atom during the strong probing process.

\begin{figure}
\includegraphics[width=8cm]{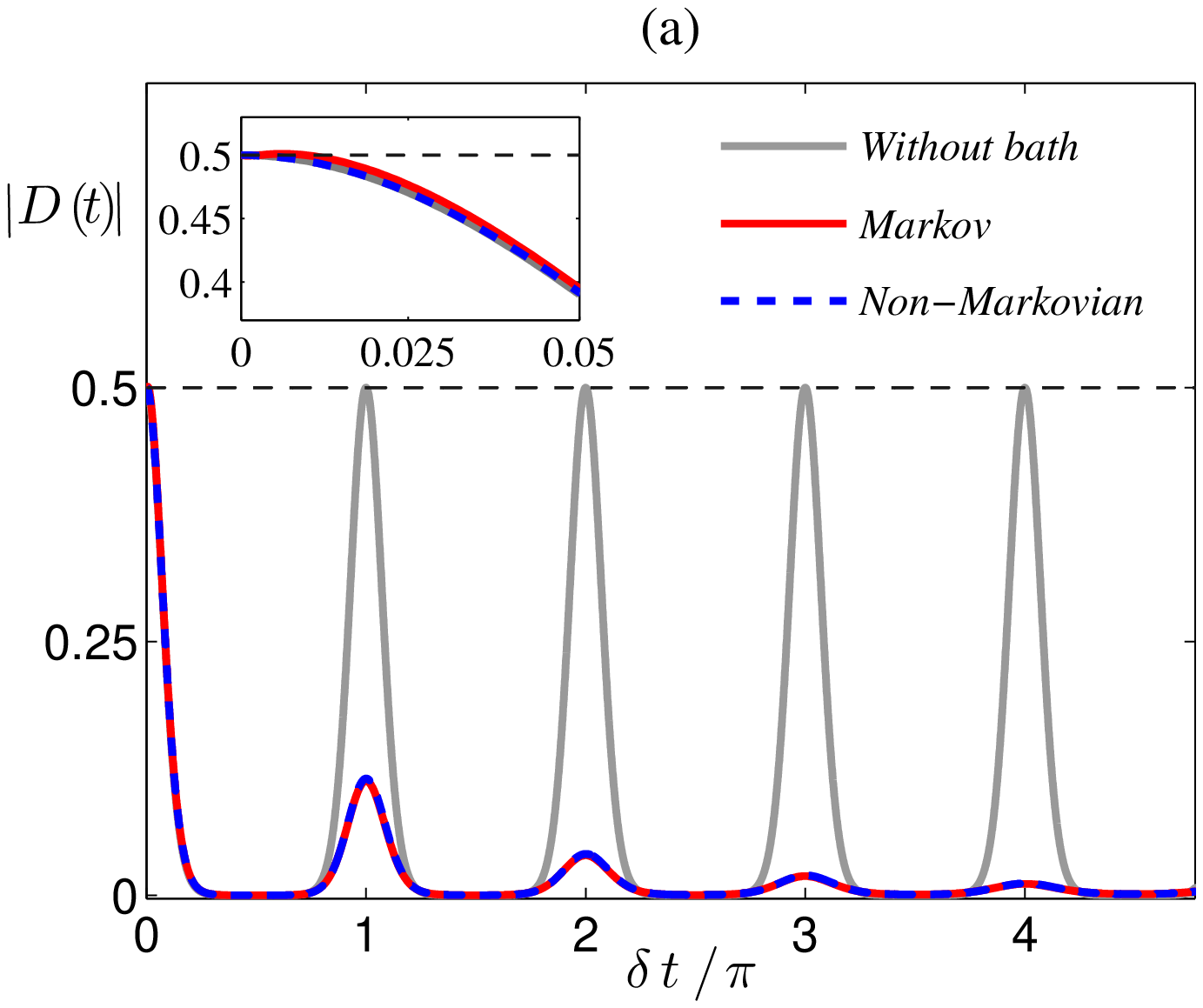}

\includegraphics[width=8cm]{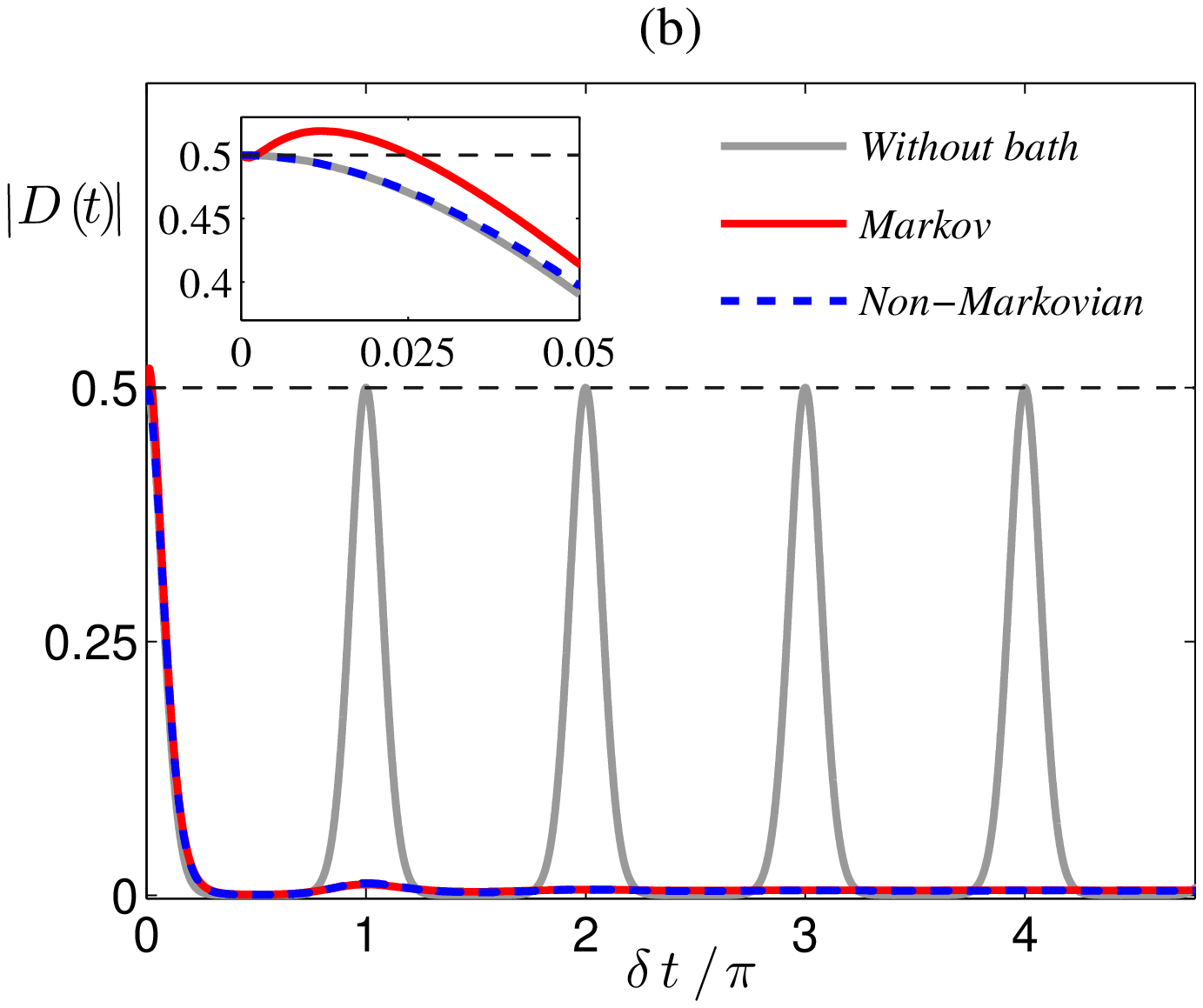}

\includegraphics[width=8cm]{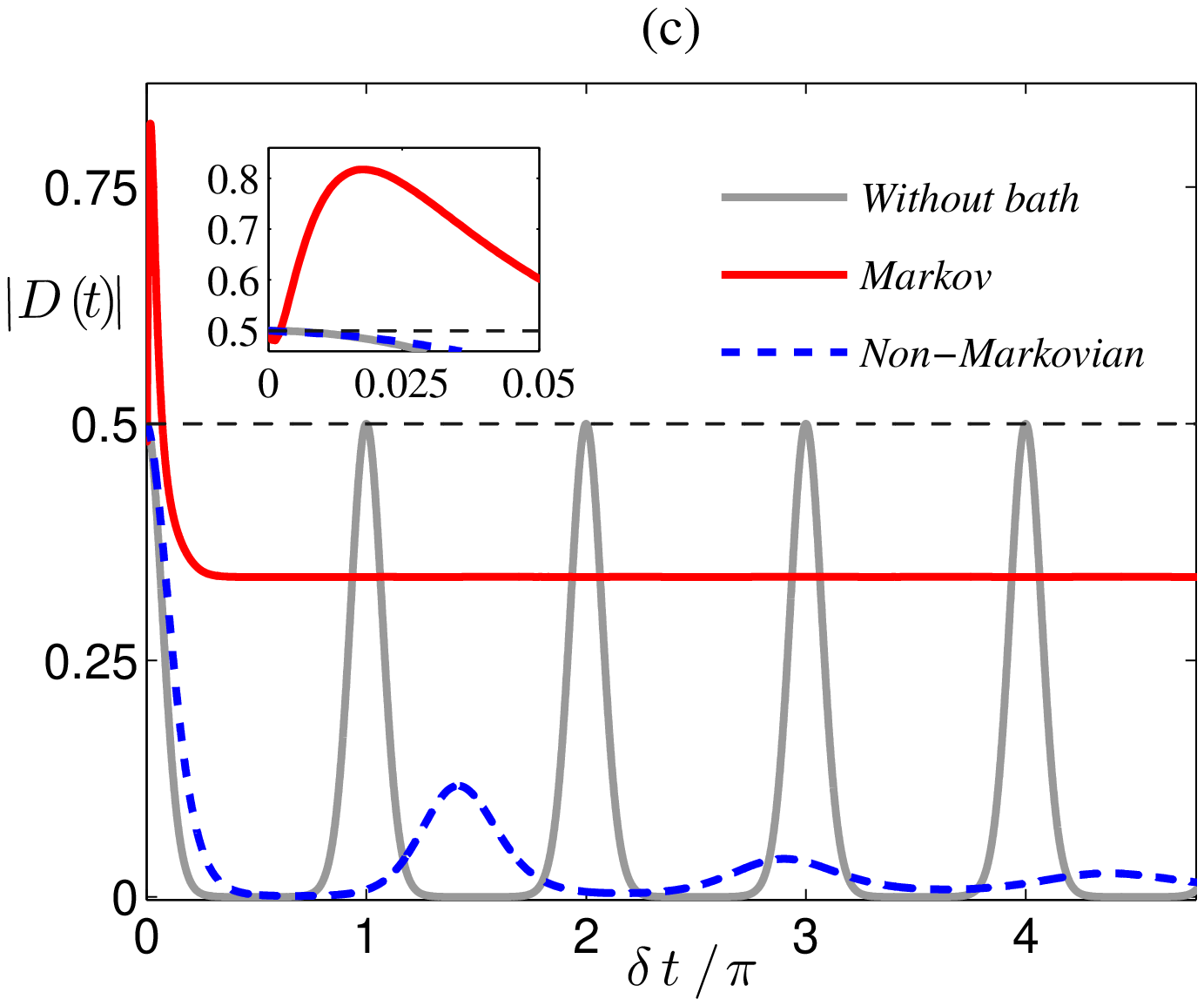}\caption{(Color online) Decoherence factor with different cavity-bath coupling
strength. ($a$) $\lambda=0.002$. ($b$) $\lambda=0.01$. ($c$)
$\lambda=0.1$.}
\end{figure}

Before entering the cavity, the atom is initialized in the superposition
state $\left(\left|e\right\rangle +\left|g\right\rangle \right)/\sqrt{2}$
and the cavity is initially in the coherent state $\left|\alpha\right\rangle $.
For simplicity, we assume that the bath is at zero temperature with
initial density matrix $\rho_{b}\left(0\right)=\left|\mathbf{0}\right\rangle \left\langle \mathbf{0}\right|$,
where $\left|\mathbf{0}\right\rangle $ is the vacuum state of the
bath. It is well known that the bath of the cavity will decrease the
coherence of the atom by disturbing the phase of the cavity field,
but it does not change the population of the atom as the result of
the dispersive atom-cavity coupling. However, we can detect this decoherence
effect by observing the Ramsey interference fringes of the out-coming
atom. The exact density matrix of the atom and field is obtained by
tracing over the degrees of freedom of the bath
\begin{eqnarray}
\!\!\!\rho_{a-f} & = & \mathrm{Tr}_{b}\left[e^{-iHt}\left(\left|\psi\left(0\right)\right\rangle \left\langle \psi\left(0\right)\right|\right)\otimes\rho_{b}\left(0\right)e^{iHt}\right],\label{eq:reduced_af}
\end{eqnarray}
where $H=H_{p}+H_{b}+H_{\mathrm{int}}$ and $\left|\psi\left(0\right)\right\rangle =\left(\left|e\right\rangle +\left|g\right\rangle \right)\otimes\left|\alpha\right\rangle /2$.
In order to describe the decoherence process of the atom, we introduce
the decoherence factor~\cite{decoh_factor}
\begin{equation}
D\left(t\right)=\frac{1}{2}e^{-\left|\alpha\right|^{2}}\mathrm{Tr}_{f}\left[\left\langle g\right|\rho_{a-f}\left|e\right\rangle \right],\label{eq:decoherence_factor}
\end{equation}
where we have added a normalization factor $\exp(-\left|\alpha\right|^{2})$.

If there were no bath present, the decoherence factor would read
\begin{equation}
D_{0}\left(t\right)=\frac{1}{2}\exp\left[\left|\alpha\right|^{2}\left(e^{-2i\delta t}-1\right)\right],
\end{equation}
which is similar to the result in Ref.~\cite{PRL1997_Haroche}. Thus
the norm of the decoherence factor will decline to a very small value
for $\left|\alpha\right|^{2}\gg1$ at the beginning and revive at
$\delta t=n\pi,(n=1,2,3,\dots)$ as depicted by the gray solid lines
in Fig.~2. Since the cavity evolves along two-pronged path in the
Hilbert space corresponding to different atomic states and the two
paths cross periodically.

When the environment of the cavity is taken into account, we obtain
the decoherence factor from Eq.~(\ref{eq:decoherence_factor})
\begin{equation}
D\left(t\right)\!=\!\frac{1}{2}\!\exp\!\left[\left(\! W_{\sigma}^{*}\!\left(t\right)\! W_{\sigma'}\!\left(t\right)\!+\! J_{3,\sigma\sigma'}\!\left(t\right)\!-\!1\!\right)\!\left|\alpha\right|^{2}\right]\!,\!\!
\end{equation}
where $W_{\sigma}$ is determined by Eq.~(\ref{eq:W_integrodiff})
with $M=\omega_{0}\pm\delta$ ($\pm$ corresponding to $\left|e\right\rangle $
and $\left|g\right\rangle $ states, respectively), and
\[
J_{3,\sigma\sigma'}\!=\!\!\int_{0}^{t}\!\!\! d\tau\!\!\!\int_{0}^{t}\!\!\! d\tau'W_{\sigma'}^{*}\!\left(\tau\right)\! W_{\sigma}\!\left(\tau'\right)\!\!\int_{0}^{\infty}\!\!\! d\omega\! J\!\left(\omega\right)\! e^{-i\omega\left(\tau-\tau'\right)}\!.
\]
Here, we choose the Ohmic spectral density with cut-off frequency
$\Omega_{c}$:
\[
J\left(\omega\right)=\lambda\omega\exp\left(-\frac{\omega}{\Omega_{c}}\right),
\]
where $\lambda$ is a dimensionless constant characterizing cavity-bath
coupling strength.

\begin{figure}
\includegraphics[width=8cm]{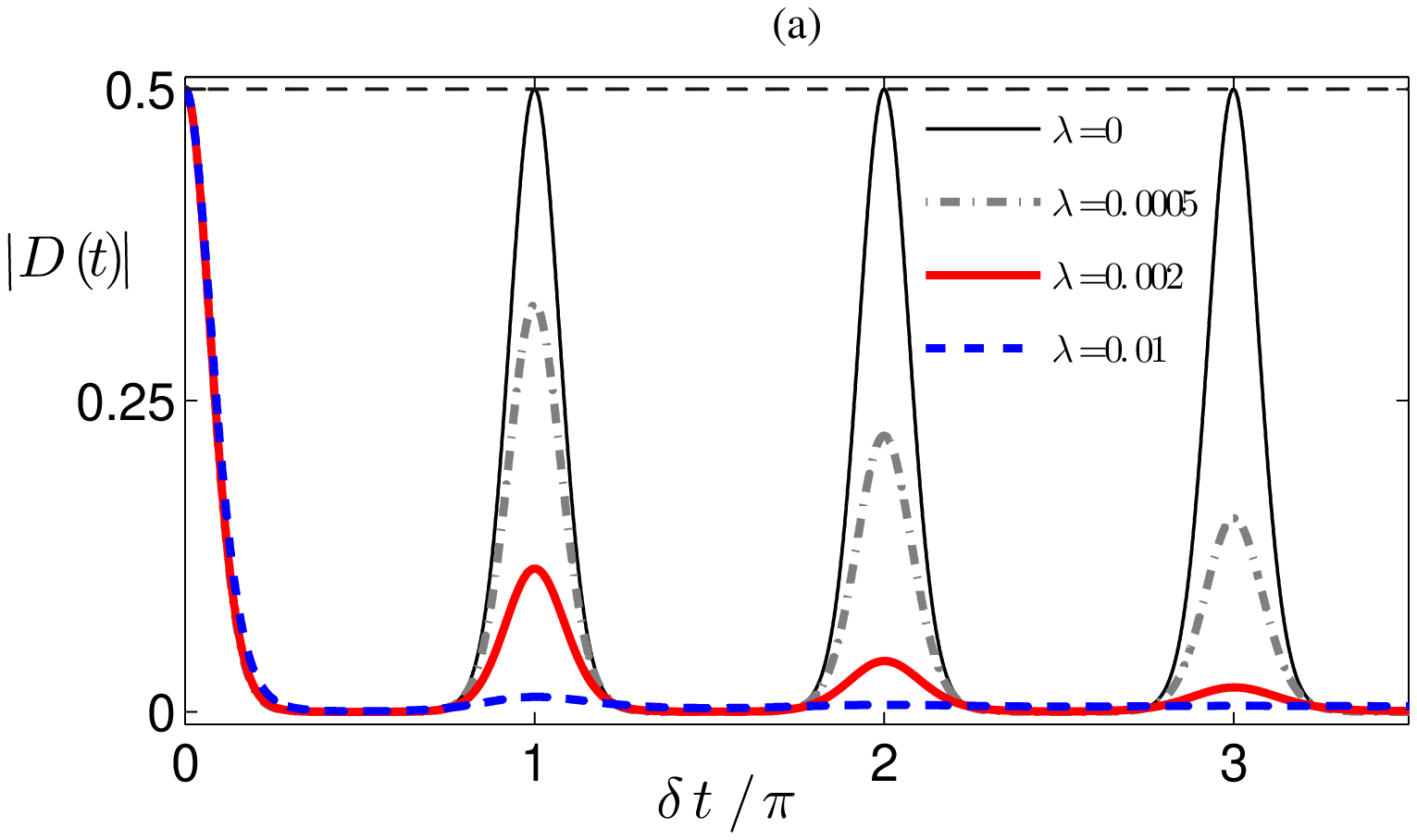}

\includegraphics[width=8cm]{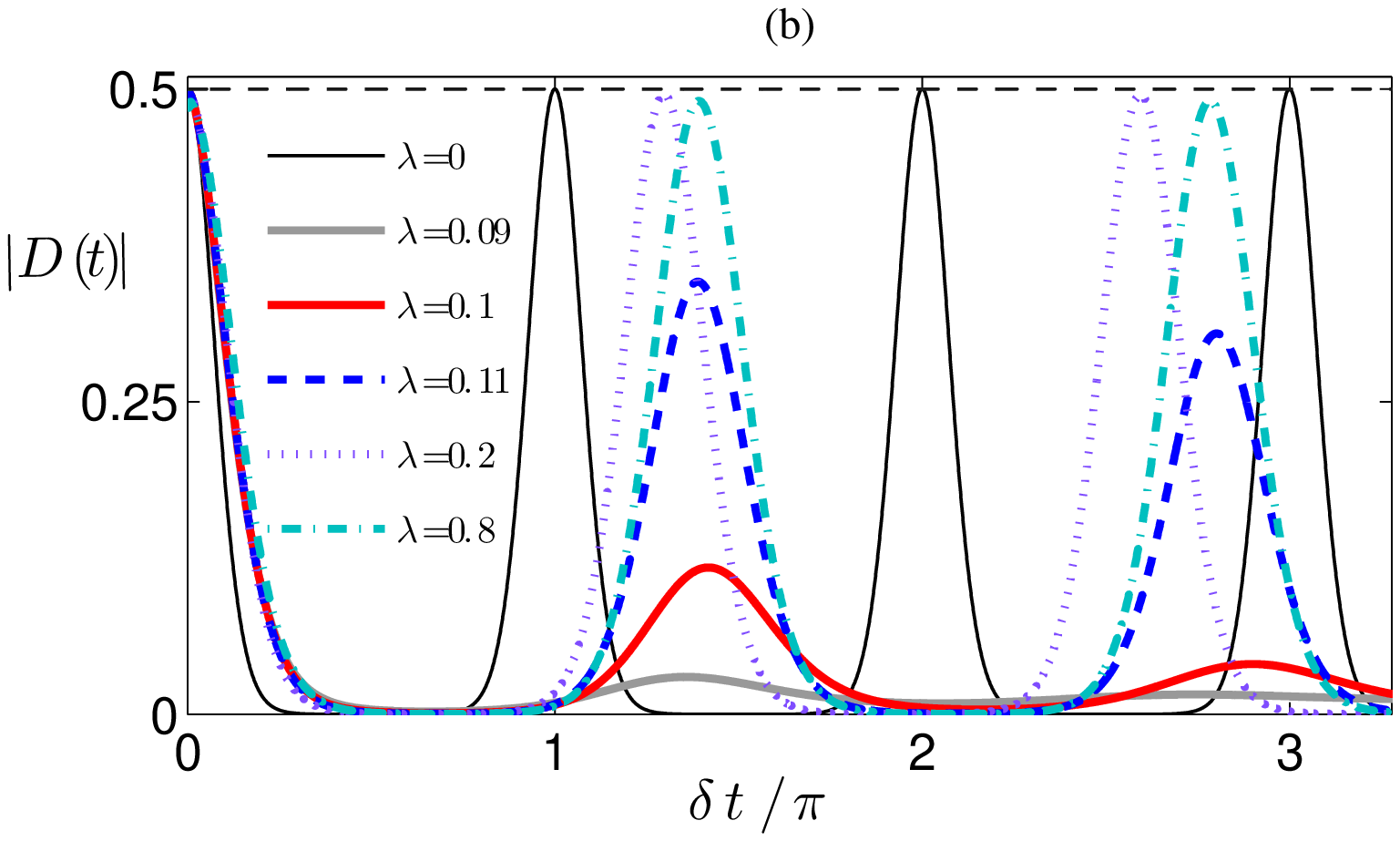}

\caption{Norm of the decoherence factor without Markov approximation. (a) If
the cavity-bath coupling strength is weak, the recovering amplitude
of the decoherence factor decreases along with $\lambda$. (a) When
the cavity-bath coupling strength is large enough, the recovering
amplitude of the decoherence factor increases along with $\lambda$
but its recovering period is changed by the bath.}
\end{figure}

Next we numerically calculate the norm of decoherence factor with
or without Markov approximation with parameters: $\omega_{0}=1$,
$\delta=0.1$, $\left|\alpha\right|^{2}=5$, and $\Omega_{c}=10$.
It is found that when the cavity-bath coupling is small ($\lambda=0.002$),
the decoherence factors with or without Markov are nearly the same
Figs.~2(a), but they diverge from each other when the coupling strength
becomes large ($\lambda=0.01$) as in Figs.~2(b). And the Markov
approximation loses its validity in strong-coupling regime ($\lambda=0.1$).
From the insets of Figs.~2(a-c), we find that the Markov approximation
also becomes invalid for a short-time dynamics (the norm of the decoherence factor
under Markov approximation exceeds $0.5$).

When the cavity-bath coupling is weak, the decoherence factor without
Markov approximation will still revive at $\delta t=n\pi,(n=1,2,3,\dots)$,
but the recovering amplitude decreases along with the cavity-bath
coupling $\lambda$ and will decay to $0$ finally (Fig.~3 (a)),
due to the dephasing of the cavity field induced by the bath. On the
contrary, if the cavity-bath coupling becomes strong enough, the reviving
magnitude will increase with the coupling strength $\lambda$~(Fig.
3(b)). Especially, when the coupling strength become to be ultra-strong
($\lambda>0.1$), the recovering amplitude almost does not decay,
just like that the bath does not exist. This is because
when $\lambda > \omega_{0}/\Omega_{c}=0.1$ (for Ohmic bath),
the cavity will stay in the system-bath coupling-induced dissipationless localized
mode~\cite{arXiv12_WMZhang}. As a result, the recovering
amplitude almost does not decay but the recovering period
is shifted.

\begin{figure}
\includegraphics[width=8cm]{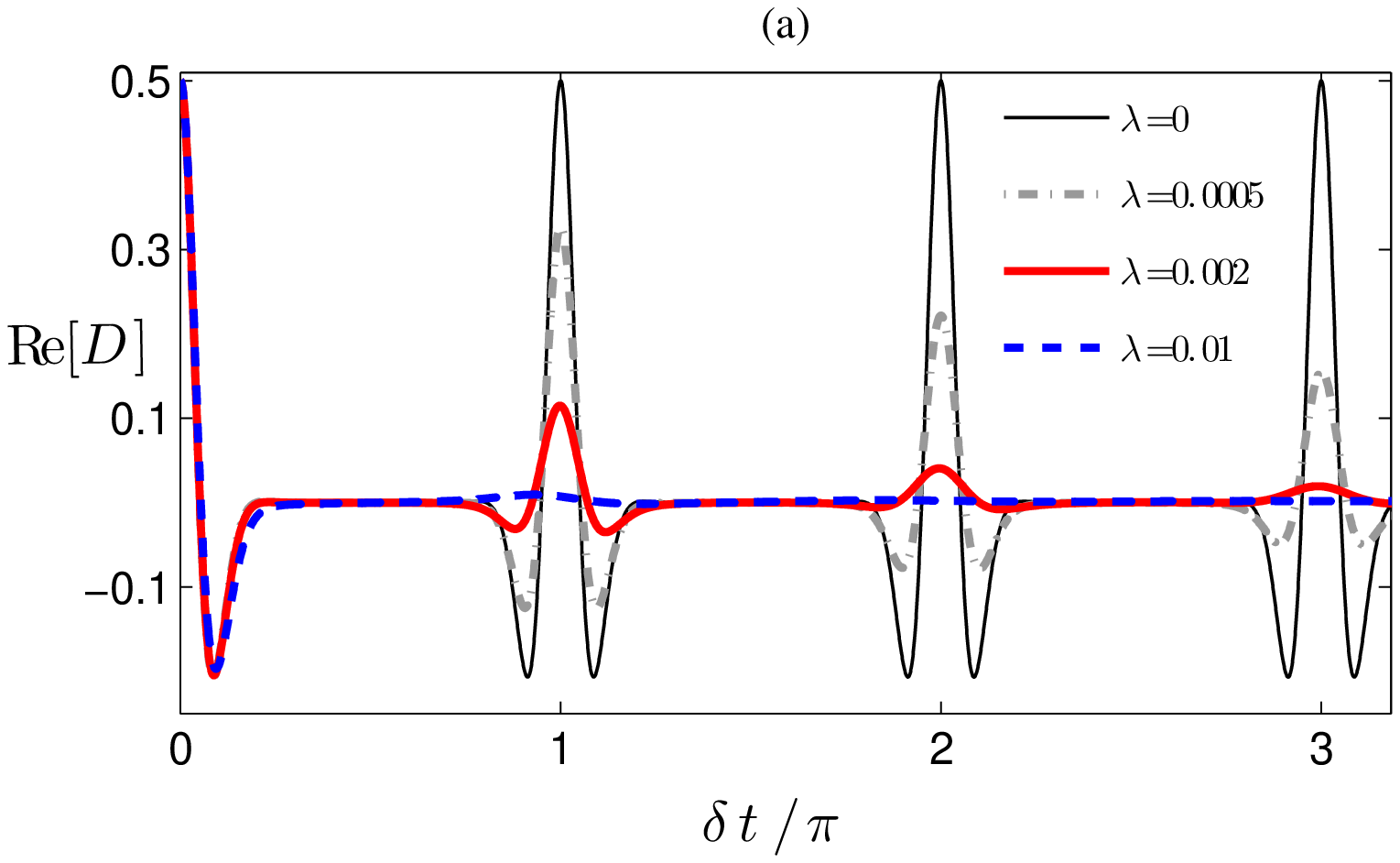}

\includegraphics[width=8cm]{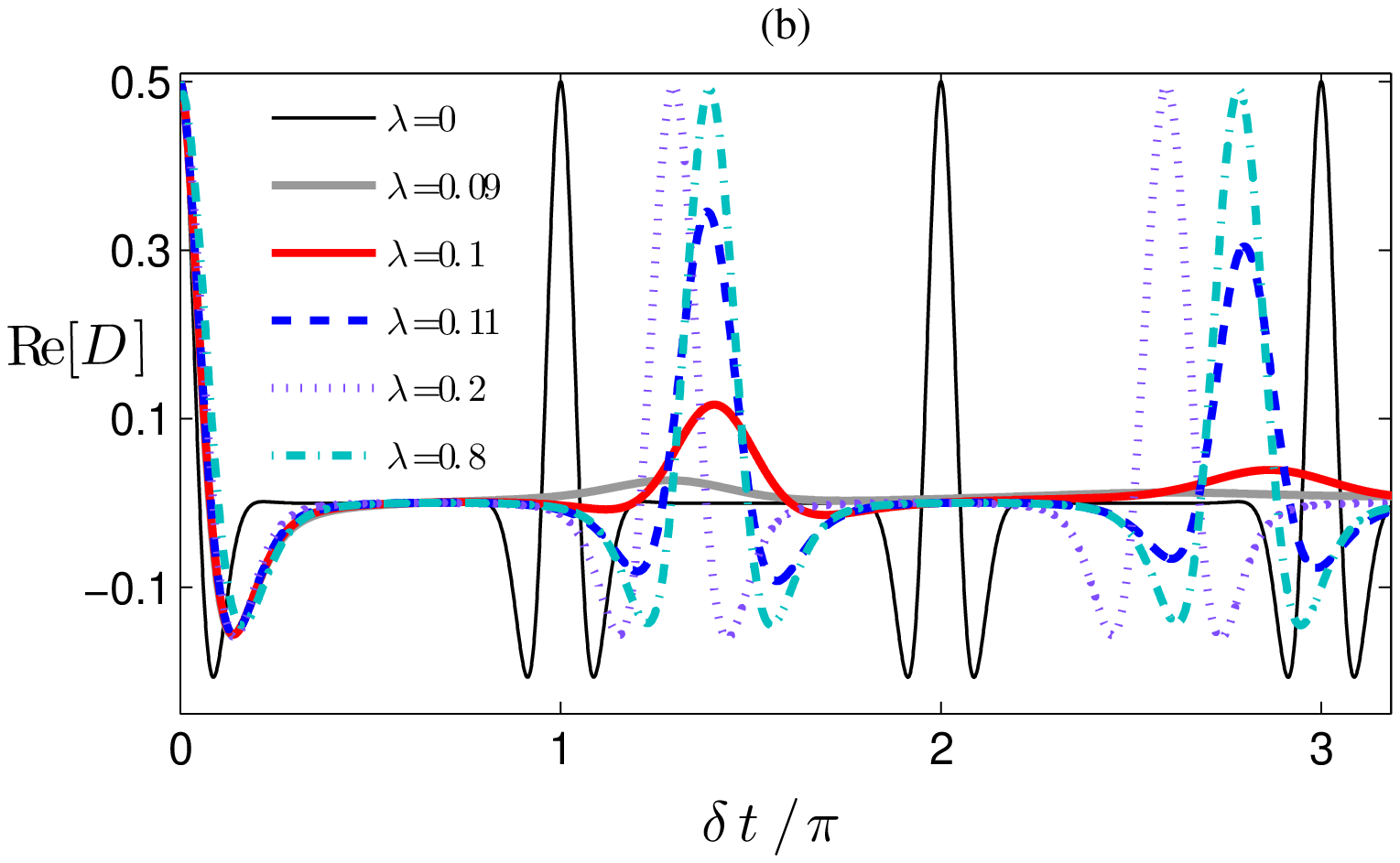}\caption{Ramsey interference is used to detect the decoherence factor. (a)
Real part of the decoherence factor in weak coupling region without
the Markov approximation. (b) Real part of the decoherence factor
in strong coupling region without the Markov approximation.}
\end{figure}

Finally, we can utilize the Ramsey interference to detect the decoherence
factor. After interacting with the cavity, the atom undergoes an additional
resonant microwave $\pi/2$ pulse performing the following transformation
\[
\left|e\right\rangle \rightarrow\frac{1}{\sqrt{2}}\left(\left|e\right\rangle +\left|g\right\rangle \right),\ \left|g\right\rangle \rightarrow\frac{1}{\sqrt{2}}\left(-\left|e\right\rangle +\left|g\right\rangle \right).
\]
And it is found that (please refer to Appendix E for the detailed
calculation)
\begin{equation}
\mathrm{Re}[D\left(t\right)]=\frac{1}{2}\left[\Pi_{g}\left(t\right)-\Pi_{e}\left(t\right)\right],\label{eq:deco_factor}
\end{equation}
where
\begin{equation}
\Pi_{\sigma}=e^{-\left|\alpha\right|^{2}}\mathrm{Tr}_{f}\left\{ \left\langle \sigma\right|e^{-i\theta\sigma_{y}/2}\rho_{a-f}e^{i\theta\sigma_{y}/2}\left|\sigma\right\rangle \right\} ,
\end{equation}
is the population of the atoms in the rotated state $\exp(i\theta\sigma_{y}/2)\left|\sigma\right\rangle (\sigma=g,e)$
with rotation angle $\theta=\pi/2$ corresponding to the final $\pi/2$
pulse. Thus we can measure the real part of the decoherence factor
through detecting the population difference of the out-coming atom.
As shown in Figs.~4, the real part of the decoherence factor can
also reflect the non-Markovianity of the bath.

\section{SUMMERY}

By constructing the reduced density matrix from the formal solution
of the Heisenberg equations, we revisited the exact non-Markovian
master equations for open quantum systems of Bose or Fermi type.
The non-Markovianity can be reflected by the time-dependent decay coefficients
such as $\Gamma\left(t\right)$ and $\tilde{\Gamma}\left(t\right)$,
with historical memory. To probe the non-Markovianity of the dissipation
of the single model EM field in a cavity, we let large detuning atoms
pass through the cavity. It displayed that the non-Markovianity of
the bath is explicitly reflected by the atomic decoherence factor.
In the week coupling regime, the periodically reviving amplitude decreases
along with the cavity-bath coupling strength $\lambda$ and decays
to $0$ finally. However, in the strong coupling regime, the reviving amplitude
increases with $\lambda$ and almost does not decay in the ultra-strong
coupling case. But the recovering period is shifted by the bath. We
expect our results to be verified by experiments.
\begin{acknowledgments}
This work is supported by National Natural Science Foundation of China
under Grants No.11121403, No. 10935010 and No. 11074261.
WMZ is supported by the National Science Council (NSC) under
Contract No. NSC-99-2112-006-008-MY3 and the National Center
for Theoretical Science of Taiwan.
\end{acknowledgments}
\appendix

\section{BOSON AND FERMION COHERENT STATES}

\subsection{Boson coherent state}

For an arbitrary complex number $\alpha=r\exp\left(i\varphi\right)$,
the coherent state of a Bose mode with frequency $\omega_{0}$ could
be defined as
\begin{equation}
\left|\alpha\right\rangle \equiv e^{\alpha a^{\dagger}}\left|0\right\rangle =\sum_{n=0}^{\infty}\frac{\alpha^{n}}{\sqrt{n!}}\left|n\right\rangle ,\label{eq:coherent_state}
\end{equation}
where $a^{\dagger}$ is the creation operator of the boson and $\left|n\right\rangle $
is the $n$th Fock state. It is found that the coherent state defined
in Eq.~(\ref{eq:coherent_state}) is not normalized and different
coherent states are generally not orthogonal
\begin{equation}
\left\langle \alpha\mid\alpha'\right\rangle =\left\langle 0\right|e^{\alpha^{*}a}e^{\alpha'a^{\dagger}}\left|0\right\rangle =\exp\left(\alpha^{*}\alpha'\right).
\end{equation}
All the coherent states form an over-complete sets
\begin{equation}
\int d\mu\left(\alpha\right)\left|\alpha\right\rangle \left\langle \alpha\right|=1,
\end{equation}
with the measures
\begin{equation}
d\mu\left(\alpha\right)\equiv e^{-\left|\alpha\right|^{2}}\frac{d^{2}\alpha}{\pi}=e^{-\left|\alpha\right|^{2}}\frac{r}{\pi}drd\varphi.
\end{equation}
And the density matrix of the thermal equilibrium state in this coherent-state
representation reads
\begin{eqnarray}
\rho_{T} & = & \frac{1}{1+f(\omega_{0})}\exp\left(-\beta\omega_{0}a^{\dagger}a\right)\\
 & = & \int d\mu\left(\alpha\right)\frac{1}{f(\omega_{0})}\exp\left[-\frac{\left|\alpha\right|^{2}}{f(\omega_{0})}\right]\left|\alpha\right\rangle \left\langle a\right|,
\end{eqnarray}
where $f\left(\omega_{0}\right)=1/\left[\exp(\beta\omega_{0})-1\right]$
is the mean occupation number, with temperature $T=1/(k_{B}\beta)$.

\subsection{Fermion coherent state}

The fermion coherent state is defined of a similar form as bosons
\begin{equation}
\left|\alpha\right\rangle \equiv e^{-\alpha a^{\dagger}}\left|0\right\rangle .\label{eq:coherent_state-2}
\end{equation}
The only difference lies in the fact that $\alpha$ is a generator
of a Grassmann algebra instead of an ordinary complex number and $a^{\dagger}$
is the creation operator for Fermi particles and they satisfy the
anti-commutation relations
\begin{equation}
\left\{ \alpha,\alpha^{\prime}\right\} =\left\{ \alpha,a\right\} =\left\{ \alpha,a^{\dagger}\right\} =0.
\end{equation}
The overlap of two fermion coherent states is
\begin{equation}
\left\langle \alpha\mid\alpha'\right\rangle =\exp\left(\alpha^{*}\alpha'\right),
\end{equation}
and the completeness relation reads
\begin{equation}
\int d\mu\left(\alpha\right)\left|\alpha\right\rangle \left\langle \alpha\right|=1,
\end{equation}
with
\begin{equation}
d\mu\left(\alpha\right)=d\alpha^{*}d\alpha e^{-\alpha^{*}\alpha}.
\end{equation}

\section{Constrains of blocks of $\mathcal{U}\left(t\right)$}

Due to hermiticity of matrix $\mathcal{H}$, the time-evolution operator
$\mathcal{U}\left(t\right)$ in Liouville space is a unitary matrix,
i.e.,
\begin{equation}
\left[\begin{array}{cc}
W\left(t\right) & T\left(t\right)\\
P\left(t\right) & Q\left(t\right)
\end{array}\right]\left[\begin{array}{cc}
W^{\dagger}\left(t\right) & P^{\dagger}\left(t\right)\\
T^{\dagger}\left(t\right) & Q^{\dagger}\left(t\right)
\end{array}\right]=I,
\end{equation}
which leads to
\begin{equation}
WW^{\dagger}+TT^{\dagger}=I,\label{eq:unitarity1}
\end{equation}
\begin{equation}
PP^{\dagger}+QQ^{\dagger}=I,\label{eq:unitarity2}
\end{equation}
\begin{equation}
WP^{\dagger}+TQ^{\dagger}=\mathbf{0},\label{eq:unitarity3}
\end{equation}
\begin{eqnarray}
PW^{\dagger}+QT^{\dagger} & = & \mathbf{0}.\label{eq:unitarity4}
\end{eqnarray}
Except some special time $t$, the matrices $W\left(t\right)$ and
$Q\left(t\right)$ are reversible. Then we have
\begin{equation}
P=-QT^{\dagger}\left(W^{\dagger}\right)^{-1},\label{eq:P}
\end{equation}
\begin{eqnarray}
P^{\dagger} & = & -W^{-1}TQ^{\dagger}.\label{eq:P_dagger}
\end{eqnarray}

\section{CALCULATION OF THE PROPAGATING FUNCTION}

The reduced density matrix $\rho_{s}\left(t\right)$ of the system
is obtained by tracing over the degrees of freedom of $B$ in $\rho\left(t\right)$
\begin{eqnarray}
\rho_{s}\left(t\right) & \!\!= & \int d\mu\!\left(\vec{z}\right)\!\left\langle \vec{z}\right|\!\rho\left(t\right)\!\left|\vec{z}\right\rangle \\
 & \equiv & \int\!\! d\mu\!\left(\vec{\alpha},\vec{\alpha}^{\prime}\right)\!\rho_{s}\left(\vec{\alpha},\vec{\alpha}^{\prime};t\right)\!\left|\vec{\alpha}\right\rangle \!\left\langle \vec{\alpha}^{\prime}\right|,
\end{eqnarray}
where $\left|\vec{\alpha}\right\rangle (\left|\vec{\alpha}^{\prime}\right\rangle )$
and $\left|\vec{z}\right\rangle $ are coherent states of $S$ and
$B$, respectively. The element of the reduced density matrix is explicitly
given by
\begin{eqnarray}
\!\!\!\!\!\!\!\!\!\!\rho_{s}\left(\vec{\alpha},\vec{\alpha}^{\prime};t\right) & = & \int d\mu\!\left(\vec{z}\right)\!\left\langle \vec{\alpha},\vec{z}\right|\!\rho\left(t\right)\!\left|\vec{\alpha}^{\prime},\vec{z}\right\rangle \\
 & \!\!\!\!\!\!\!\!\!\!\!= & \!\!\!\int\!\! d\mu\!\left(\vec{z}\right)\! d\mu\!\left(\!\vec{\xi},\!\vec{z}^{\prime}\!,\vec{\xi}^{\prime}\!,\!\vec{z}^{\prime\prime}\!\right)\left\langle \vec{\alpha},\vec{z}\right|\! U\!\left(t\right)\!\left|\vec{\xi},\vec{z}^{\prime}\right\rangle \nonumber \\
 &  & \!\!\!\!\!\!\!\!\!\!\!\!\!\times\!\left\langle \!\vec{\xi},\vec{z}^{\prime}\right|\!\rho_{s}\!\left(0\right)\!\rho_{b}\!\left(0\right)\!\left|\!\vec{\xi}^{\prime},\vec{z}^{\prime\prime}\!\right\rangle \!\!\left\langle \!\vec{\xi}^{\prime},\vec{z}^{\prime\prime}\right|\! U^{\dagger}\!\left|\vec{\alpha}^{\prime},\vec{z}\right\rangle \\
 & \!\!\!\!\!\!\!\!\!\!\!\!\!\!\!\equiv & \!\!\!\!\!\!\!\!\!\int\!\!\! d\mu\!\left(\!\vec{\xi},\vec{\xi}^{\prime}\!\right)\! K\!\left(\vec{\alpha}^{\dagger},\vec{\alpha}^{\prime},\vec{\xi},\vec{\xi}^{\prime\dagger},t\right)\!\!\left\langle \vec{\xi}\right|\!\rho_{s}\!\left(0\right)\!\left|\vec{\xi}^{\prime}\right\rangle \!\!,
\end{eqnarray}
with
\begin{eqnarray}
K & = & \int d\mu\!\left(\vec{z}\right)\left\langle \vec{\alpha},\vec{z}\right|U\left(t\right)\left|\vec{\xi}\right\rangle \nonumber \\
 &  & \times\!\left\langle \vec{\xi}^{\prime}\right|\rho_{b}\left(0\right)U^{\dagger}\left(t\right)\left|\vec{\alpha}^{\prime},\vec{z}\right\rangle .
\end{eqnarray}
Here, we have used the fact the initial state of the total system
is of the direct product form and the completeness of the coherent
states $\left\{ \left|\vec{z}^{\prime}\right\rangle \right\} $ and
$\left\{ \left|\vec{z}^{\prime\prime}\right\rangle \right\} $ of
the bath.

With the help of Eqs.~(\ref{eq:solu1}), (\ref{eq:solu2}), (\ref{eq:wave_function}),
(\ref{eq:propagating_f}), and (\ref{eq:ro_b}), the propagator is
re-expressed in terms of the coefficient matrices
\begin{eqnarray}
 &  & K\left(\vec{\alpha}^{\dagger},\vec{\alpha}^{\prime},\vec{\xi},\vec{\xi}^{\prime\dagger};t\right)\nonumber \\
 & = & \int d\mu\left(\vec{z}\right)\left\langle W^{\dagger}\vec{\alpha}+P^{\dagger}\vec{z}\mid\vec{\xi}\right\rangle \left\langle \vec{\xi}^{\prime}\mid W^{\dagger}\vec{\alpha}^{\prime}+P^{\dagger}\vec{z}\right\rangle \nonumber \\
\!\!\!\!\!\! &  & \!\!\!\!\!\!\!\!\!\times\!\left(\!\!\prod_{l}\!\frac{1}{f_{l}\!+\!1}\!\!\right)\!\left\langle T^{\dagger}\!\vec{\alpha}\!+\! Q^{\dagger}\vec{z}\right|\!\exp\!\!\left[\!-\!\vec{b}^{\dagger}\!\beta E\vec{b}\right]\!\!\left|T^{\dagger}\!\vec{\alpha}^{\prime}\!+\! Q^{\dagger}\!\vec{z}\right\rangle \!\!.
\end{eqnarray}
 Using formulas $\left\langle \alpha\right|\exp\left(\delta b^{\dagger}b\right)\left|\alpha'\right\rangle =\exp\left[\alpha^{*}\alpha'\exp\left(\delta\right)\right]$
and
\begin{equation}
\int d\mu\left(\vec{z}\right)e^{\vec{z}^{\dagger}D\vec{z}+\vec{u}^{\dagger}\cdot\vec{z}+\vec{z}^{\dagger}\cdot\vec{v}}=\frac{\exp\left[\vec{u}^{\dagger}\left(I-D\right)^{-1}\vec{v}\right]}{\mathrm{det}\left[I-D\right]},
\end{equation}
(for any $N_{s}\times N_{s}$ Hermitian matrix $D$ makes $\left(I-D\right)$
positive-definite), one goes to
\begin{eqnarray}
\!\!\!\!\!\!\!\!\!\!\!\!\!K(\vec{\alpha}^{\dagger},\vec{\alpha}^{\prime},\vec{\xi},\vec{\xi}^{\prime\dagger},t)\! & = & \!A(t)\!\exp\!\!\left[\!\!\begin{array}{c}
\vec{\alpha}^{\dagger}J_{1}(t)\vec{\xi}+\vec{\xi}^{\prime\dagger}J_{1}^{\dagger}(t)\vec{\alpha}^{\prime}\\
+\vec{\alpha}^{\dagger}J_{2}(t)\vec{\alpha}^{\prime}+\vec{\xi}^{\prime\dagger}J_{3}\vec{\xi}\!\!\!\!
\end{array}\right]\!,
\end{eqnarray}
where
\begin{equation}
A=\left(\prod_{l}\frac{1}{f_{l}\!+\!1}\right)\mathrm{det}\left[I-Qf\left(I+f\right)^{-1}Q^{\dagger}\right]^{-1},\label{eq:A}
\end{equation}
\begin{equation}
\!\! J_{1}\!=W\!+\! Tf\!\left(I\!+\! f\right)^{-1}\! Q^{\dagger}\!\!\left[I\!-\! Qf\!\left(I\!+\! f\right)^{-1}\! Q^{\dagger}\right]^{-1}\!\!\! P\!\!,\label{eq:J_1}
\end{equation}
\begin{eqnarray}
\!\!\!\!\!\!\!\!\!\!\!\! J_{2} & = & Tf\!\left(I\!+\! f\right)^{-1}T^{\dagger}+\! T\! f\!\left(I\!+\! f\right)^{-1}\! Q^{\dagger}\nonumber \\
 &  & \times\left[I-Qf\left(I\!+\! f\right)^{-1}\! Q^{\dagger}\right]^{-1}\! Q\! f\!\left(I\!+\! f\right)^{-1}T^{\dagger},\label{eq:J_2}
\end{eqnarray}
\begin{eqnarray}
J_{3} & = & P^{\dagger}\left[I-Qf\left(I\!+\! f\right)^{-1}\! Q^{\dagger}\right]^{-1}\! P,\label{eq:J_3}
\end{eqnarray}
and we have introduced a diagonal matrix $f=\mathrm{diag}[f_{1},f_{2},\cdots,f_{N_{b}}]$.

Then we will deal with these four terms one by one. First we make
some pretreatment to obtain an expanding series. From Eqs.~(\ref{eq:unitarity2}),
(\ref{eq:P}), and (\ref{eq:P_dagger}), one finds
\begin{equation}
\! I\!-\! Q\! f\!\left(I\!+\! f\right)^{-1}\!\! Q^{\dagger}\!\!=\!\! Q\!\!\left[T^{\dagger}\!\left(WW^{\dagger}\right)^{-1}\!\! T\!+\!\left(\! I\!+\! f\right)^{-1}\right]\! Q^{\dagger}\!\!.\!\!\!\!\!
\end{equation}
So that
\begin{eqnarray}
\!\!\!\!\!\!\!\!\!\!\!\!\!\!\!\!\!\!\! &  & \left[I-Qf\left(I\!+\! f\right)^{-1}\! Q^{\dagger}\right]^{-1}\nonumber \\
\!\!\!\!\!\!\!\!\!\!\!\!\!\!\!\!\!\!\! & =\!\! & \left(Q^{\dagger}\right)^{-1}\!\!\left(\! I\!+\! f\right)\!\!\sum_{n=0}^{\infty}\!\left(-1\right)^{n}\!\!\left[\! T^{\dagger}\!\!\left(WW^{\dagger}\right)^{-1}\! T\!\left(I\!+\! f\right)\right]^{n}\!\! Q^{-1}\!\!\!.\label{eq:core}
\end{eqnarray}

\subsection{$J_{1}\left(t\right)$, $J_{2}\left(t\right)$, and $J_{3}\left(t\right)$}

According to Eqs.~(\ref{eq:P}), (\ref{eq:J_1}), and (\ref{eq:core}),
$J_{1}\left(t\right)$ is explicitly expanded to
\begin{eqnarray}
J_{1} & = & W-Tf\sum_{n=0}^{\infty}\left(-1\right)^{n}\!\!\left[\! T^{\dagger}\!\!\left(WW^{\dagger}\right)^{-1}\! T\!\left(I\!+\! f\right)\right]^{n}T^{\dagger}\left(W^{\dagger}\right)^{-1}\nonumber \\
 & = & W-TfT^{\dagger}\sum_{n=0}^{\infty}\left(-1\right)^{n}\!\!\left[\!\!\left(WW^{\dagger}\right)^{-1}\! T\!\left(I\!+\! f\right)T^{\dagger}\!\right]^{n}\left(W^{\dagger}\right)^{-1}\nonumber \\
 & = & W-V\left[1+\left(WW^{\dagger}\right)^{-1}\! T\!\left(I\!+\! f\right)T^{\dagger}\right]^{-1}\left(W^{\dagger}\right)^{-1}\nonumber \\
 & = & W-V\left[WW^{\dagger}+T\!\left(I\!+\! f\right)T^{\dagger}\right]^{-1}W\\
 & = & \left(1+V\right)^{-1}W.
\end{eqnarray}
The third step we have introduced a new $N_{s}\times N_{s}$-matrix
$V\left(t\right)=T\left(t\right)fT^{\dagger}\left(t\right)$. Similarly,
one obtains
\begin{equation}
J_{3}=I-W^{\dagger}\left(1+V\right)^{-1}W.\label{eq:J_3final}
\end{equation}
The calculation of $J_{2}$ is a little more complicated
\begin{eqnarray}
\!\!\!\!\!\!\!\!\! J_{2} & = & T\left\{ I+\! f\!\sum_{n=0}^{\infty}\left(-1\right)^{n}\!\!\left[\! T^{\dagger}\!\!\left(WW^{\dagger}\right)^{-1}\! T\!\left(I\!+\! f\right)\right]^{n}\right\} \nonumber \\
 &  & \times f\!\left(I\!+\! f\right)^{-1}T^{\dagger}\\
 & = & V+Tf\left(\sum_{n=1}^{\infty}\left(-1\right)^{n}\!\!\left[\! T^{\dagger}\!\!\left(WW^{\dagger}\right)^{-1}\! T\!\left(I\!+\! f\right)\right]^{n-1}\right)\nonumber \\
 &  & \times T^{\dagger}\left(WW^{\dagger}\right)^{-1}TfT^{\dagger}\\
 & = & V+Tf\left(\sum_{n=1}^{\infty}\left(-1\right)^{n}\!\!\left[\! T^{\dagger}\!\!\left(WW^{\dagger}\right)^{-1}\! T\!\left(I\!+\! f\right)\right]^{n-1}\right)\nonumber \\
 &  & \times T^{\dagger}\left(WW^{\dagger}\right)^{-1}T\left(I+f-I\right)T^{\dagger}\\
 & = & V\left\{ I+\sum_{n=1}^{\infty}\left(-1\right)^{n}\!\!\left[\!\left(WW^{\dagger}\right)^{-1}\! T\!\left(I\!+\! f\right)\! T^{\dagger}\!\right]^{n}\right\} \nonumber \\
 &  & \!\!\!\!\!\!+V\!\!\sum_{n=0}^{\infty}\!\left(\!-1\right)^{n}\!\!\left[\!\left(W\! W^{\dagger}\!\right)^{-1}\! T\!\left(I\!+\! f\right)\! T^{\dagger}\!\right]^{n}\!\!\!\left(\! W\! W^{\dagger}\right)^{\!-1}\!\!\! TT^{\dagger}\\
 & = & V\left[WW^{\dagger}+T\left(I+f\right)T^{\dagger}\right]^{-1}\left(WW^{\dagger}+TT^{\dagger}\right)\\
 & = & V\left(I+V\right)^{-1}.
\end{eqnarray}

\subsection{$A\left(t\right)$}

The matrix $A\left(t\right)$ is determined by the normalization condition,
\begin{eqnarray}
\!\!\!1 & = & \mathrm{Tr}\!\left[\rho_{s}\!\left(\! t\!\right)\right]\\
 & = & \int\!\!\! d\!\!\mu\!\left(\vec{\alpha}\right)\!\! d\!\mu\!\!\left(\vec{\xi},\vec{\xi}^{\prime}\right)\!\! K\!\!\left(\!\vec{\alpha},\!\vec{\alpha},\!\vec{\xi},\!\vec{\xi}^{\prime}\!,t\right)\!\!\left\langle \!\vec{\xi}\right|\!\rho_{s}\!\left(0\right)\!\left|\!\vec{\xi}^{\prime}\right\rangle \\
 & = & A\mathrm{det}\!\left[I\!+\! V\right]\!\!\!\int\!\! d\!\mu\!\!\left(\!\vec{\xi},\vec{\xi}^{\prime}\right)\!\!\left\langle \!\vec{\xi}\right|\!\rho_{s}\!\left(0\right)\!\left|\!\vec{\xi}^{\prime}\right\rangle \!\exp\!\left(\!\vec{\xi}^{\prime\dagger}\!\cdot\!\vec{\xi}\right)\\
 & = & A\mathrm{det}\left[I+V\right]\int\!\! d\!\mu\!\!\left(\vec{\xi}\right)\!\!\left\langle \!\vec{\xi}\right|\!\rho_{s}\!\left(0\right)\!\left|\!\vec{\xi}\right\rangle .
\end{eqnarray}
In the second step, we carried out the integral over $\vec{\alpha}$
of Eq. (\ref{eq:K}) and used the identity
\begin{equation}
J_{3}\left(t\right)+J_{1}^{\dagger}\left(I+V\right)J_{1}\left(t\right)=I.
\end{equation}
And the last step, the following formula is used
\begin{equation}
\int d\mu\left(\alpha^{\prime}\right)\left(\alpha^{\prime}\right)^{n}e^{\alpha^{\prime*}\alpha}=\alpha^{n}.
\end{equation}
Since the initial density matrix is also normalized, thus
\[
A\left(t\right)=\mathrm{det}\left[\left(I+V\right)^{-1}\right].
\]

\section{TIME DIFFERENTIAL OF THE PROPAGATING FUNCTION}

The time differential of the propagating function is given by
\begin{equation}
\dot{K}=\left[\frac{\dot{A}}{A}\!+\!\vec{\alpha}^{\dagger}\dot{J}_{1}\vec{\xi}\!+\!\vec{\xi}^{\prime\dagger}\dot{J}_{1}^{\dagger}\vec{\alpha}^{\prime}\!+\!\vec{\alpha}^{\dagger}\dot{J}_{2}\vec{\alpha}^{\prime}\!+\!\vec{\xi}^{\prime\dagger}\dot{J}_{3}\vec{\xi}\right]K.
\end{equation}
 We define the differential operators
\begin{equation}
\overset{\rightarrow}{\nabla}_{\alpha^{*}}\equiv\left[\frac{\partial}{\partial\alpha_{1}^{*}},\frac{\partial}{\partial\alpha_{2}^{*}},\cdots,\frac{\partial}{\partial\alpha_{N_{s}}^{*}}\right]^{T},
\end{equation}
and
\begin{eqnarray}
\overset{\rightarrow}{\nabla}_{\alpha}^{T} & \equiv & \left[\frac{\partial}{\partial\alpha_{1}},\frac{\partial}{\partial\alpha_{2}},\cdots,\frac{\partial}{\partial\alpha_{N_{s}}}\right].
\end{eqnarray}
It is ready to find that
\begin{equation}
\vec{\xi}K=J_{1}^{-1}\left(\overset{\rightarrow}{\nabla}_{\alpha^{*}}-J_{2}\vec{\alpha}^{\prime}\right)K,
\end{equation}
\begin{equation}
\vec{\xi}^{\prime\dagger}K=\left(\overset{\rightarrow}{\nabla}_{\alpha'}^{T}-\vec{\alpha}^{\dagger}J_{2}\right)\left(J_{1}^{\dagger}\right)^{-1}K,
\end{equation}
\begin{eqnarray}
\!\!\!\!\!\!\!\!\vec{\xi}^{\prime\dagger}\! K\!\vec{\xi} & \!\!=\!\!\! & \left(\!\overset{\rightarrow}{\nabla}_{\alpha'}^{T}\!-\!\vec{\alpha}^{\dagger}\! J_{2}\!\!\right)\!\left(\! J_{1}^{\dagger}\!\right)^{-1}\!\! J_{1}^{-1}\!\!\left(\!\overset{\rightarrow}{\nabla}_{\alpha^{*}}\!-\! J_{2}\vec{\alpha}^{\prime}\!\right)\! K.
\end{eqnarray}
These relations lead to
\begin{eqnarray}
\dot{K}\!\! & =\!\! & \vec{\alpha}^{\dagger}\!\!\left[\!\dot{J}_{2}\!-\!\dot{J}_{1}\! J_{1}^{-1}J_{2}\!\!-\! J_{2}\!\left(\! J_{1}^{\dagger}\!\right)^{-1}\!\!\dot{J}_{1}^{\dagger}\!+\! J_{2}\!\left(\! J_{1}^{\dagger}\!\right)^{-1}\!\!\dot{J}_{3}J_{1}^{-1}\! J_{2}\!\right]\!\! K\!\vec{\alpha}^{\prime}\nonumber \\
 &  & +\left\{ \frac{\dot{A}}{A}-\mathrm{Tr}\left[\left(J_{1}^{\dagger}\right)^{-1}\dot{J}_{3}J_{1}^{-1}J_{2}\right]\right\} K\nonumber \\
 &  & +\vec{\alpha}^{\dagger}\left[\!\dot{J}_{1}\! J_{1}^{-1}-\! J_{2}\!\left(\! J_{1}^{\dagger}\!\right)^{-1}\!\!\dot{J}_{3}J_{1}^{-1}\!\right]\overset{\rightarrow}{\nabla}_{\alpha^{*}}K\nonumber \\
 &  & +\left(\overset{\rightarrow}{\nabla}_{\alpha'}^{T}\! K\right)\left[\!\left(\! J_{1}^{\dagger}\!\right)^{-1}\!\!\dot{J}_{1}^{\dagger}-\left(\! J_{1}^{\dagger}\!\right)^{-1}\!\!\dot{J}_{3}J_{1}^{-1}\! J_{2}\!\right]\vec{\alpha}^{\prime}\nonumber \\
 &  & +\overset{\rightarrow}{\nabla}_{\alpha'}^{T}\!\left(\! J_{1}^{\dagger}\!\right)^{-1}\!\!\dot{J}_{3}J_{1}^{-1}\overset{\rightarrow}{\nabla}_{\alpha^{*}}K\\
 & \equiv & \vec{\alpha}^{\dagger}\tilde{\Gamma}K\vec{\alpha}^{\prime}-\mathrm{Tr}\left[\tilde{\Gamma}\right]K-\vec{\alpha}^{\dagger}\left(\Gamma+\mathrm{i}\tilde{\Omega}+\tilde{\Gamma}\right)\overset{\rightarrow}{\nabla}_{\alpha^{*}}K\nonumber \\
\!\!\!\!\!\! & \!\!\!\!\!\! & -\!\left(\!\overset{\rightarrow}{\nabla}_{\alpha'}^{T}\! K\!\right)\!\!\left(\Gamma\!-\mathrm{i}\tilde{\Omega}\!+\!\tilde{\Gamma}\!\right)\!\vec{\alpha}^{\prime}\!+\!\overset{\rightarrow}{\nabla}_{\alpha'}^{T}\!\left(\tilde{\Gamma}\!+\!2\Gamma\!\right)\!\overset{\rightarrow}{\nabla}_{\alpha^{*}}\! K\!,
\end{eqnarray}
with Hermitian matrices
\begin{equation}
\tilde{\Gamma}=\dot{V}-\dot{W}W^{-1}V-V\left(\dot{W}W^{-1}\right)^{\dagger},
\end{equation}
\begin{equation}
\Gamma=-\frac{1}{2}\left[\dot{W}W^{-1}+\left(\dot{W}W^{-1}\right)^{\dagger}\right],
\end{equation}
and
\begin{equation}
\tilde{\Omega}=\frac{\mathrm{i}}{2}\left[\dot{W}W^{-1}-\left(\dot{W}W^{-1}\right)^{\dagger}\right].
\end{equation}
The last step the following relations have been used
\begin{eqnarray}
\!\!\!\!\!\!\!\!\!\!\!\!\!\!\!\!\!\!\!\dot{J}_{1}\! J_{1}^{-1}\!\! & = & \!\!\left[\!\frac{d}{dt}\!\!\left(I\!+\! V\right)^{-1}\!\right]\!\!\!\left(I\!+\! V\right)\!\!+\!\!\left(I\!+\! V\right)^{-1}\!\!\left(\!\dot{W}W^{-1}\!\right)\!\!\left(I\!\!+\!\! V\right)\\
\!\!\!\!\!\!\!\!\!\!\!\!\!\!\!\! & = & -\left[\left(I+V\right)^{-1}\dot{V}\left(I+V\right)^{-1}\right]\left(I+V\right)\nonumber \\
\!\!\!\!\!\!\!\!\!\!\!\!\!\!\!\! &  & +\left(I+V\right)^{-1}\left(\dot{W}W^{-1}\right)\left(I+V\right)\\
\!\!\!\!\!\!\!\!\!\!\!\!\!\!\!\! & = & \!\!-\!\left(I\!+\! V\right)^{-1}\!\!\left[\dot{V}-\!\!\left(\!\dot{W}W^{-1}\!\right)\!\!\left(I\!+V\right)\!\right]\!\!,
\end{eqnarray}
\begin{equation}
\left(\! J_{1}^{\dagger}\!\right)^{-1}\!\!\!\dot{J}_{3}J_{1}^{-1}\!\!=\!\!-\!\left(I\!\!+\!\! V\!\right)\!\!\left(\!\dot{W}\! W^{-1}\!\right)^{\dagger}\!\!\!-\!\dot{W}\! W^{-1}\!\!\left(I\!\!+\!\! V\right)\!+\!\dot{V}\!,
\end{equation}
and
\begin{eqnarray}
\!\!\!\!\!\! &  & \frac{\dot{A}}{A}-\mathrm{Tr}\left[\left(J_{1}^{\dagger}\right)^{-1}\dot{J}_{3}J_{1}^{-1}J_{2}\right]\nonumber \\
 & =\!\! & \frac{d}{dt}\mathrm{ln}A\!+\!\mathrm{Tr}\!\left[\begin{array}{c}
\!\! V\left(\dot{W}W^{-1}\!\right)^{\dagger}\!+\!\left(\!\dot{W}W^{-1}\!\right)\!\! V\!\!\\
-\dot{V}\left[I-\left(I+V\right)^{-1}\right]
\end{array}\right]\\
 & = & \!\!\frac{d}{dt}\mathrm{Tr}\!\left[\mathrm{ln}\left(I\!+\! V\right)^{-1}\right]\!+\!\mathrm{Tr}\!\left[-\tilde{\Gamma}\!+\!\dot{V}\!\left(I\!+\! V\right)^{-1}\!\right]\\
 & = & -\mathrm{Tr}\left[\tilde{\Gamma}\right]
\end{eqnarray}

\section{DECOHERENCE FACTOR}

Through the approach in Appendix C, we can obtain the element of the
reduced density in Eq~(\ref{eq:reduced_af})
\begin{eqnarray}
 &  & \rho_{a-f}\left(\!\alpha_{f},\sigma;\alpha_{f}^{\prime},\sigma'\!\right)\nonumber \\
 & = & \frac{1}{2}A_{\sigma\sigma'}\exp\left[\begin{array}{c}
\alpha_{f}^{*}J_{1,\sigma\sigma'}\alpha+\alpha^{*}J_{1,\sigma'\sigma}^{\dagger}\alpha_{f}^{\prime}\\
+\alpha_{f}^{*}J_{2,\sigma\sigma'}\alpha_{f}^{\prime}+\alpha^{*}J_{3,\sigma\sigma'}\alpha
\end{array}\right],\label{eq:ele_red_den}
\end{eqnarray}
where in the case of zero temperature bath
\begin{equation}
\! A_{\sigma\sigma'}\!=\!1,J_{1,\sigma\sigma'}\!=\! W_{\sigma}\left(t\right),J_{2,\sigma\sigma'}\!=\!0,J_{3,\sigma\sigma'}\!=\! P_{\sigma'}^{\dagger}P_{\sigma}.\!\!\!\!
\end{equation}
Here $W_{\sigma}$ is determined by Eq.~(\ref{eq:W_integrodiff})
with $M=\omega_{0}\pm\delta$ ($\pm$ corresponding to $\left|e\right\rangle $
and $\left|g\right\rangle $ states, respectively) and $P_{\sigma}$
is given by Eq.~(\ref{eq:P_W}). Following from Eq.~(\ref{eq:decoherence_factor}),
we find that the population difference of the out-coming atom just
gives the decoherence factor
\begin{eqnarray}
 &  & \Pi_{g}\left(t\right)-\Pi_{e}\left(t\right)\nonumber \\
 & = & e^{-\left|\alpha\right|^{2}}\mathrm{Tr}_{f}\left\{ \begin{array}{c}
\left\langle g\right|e^{-i\theta\sigma_{y}/2}\rho_{a-f}e^{i\theta\sigma_{y}/2}\left|g\right\rangle \\
-\left\langle e\right|e^{-i\theta\sigma_{y}/2}\rho_{a-f}e^{i\theta\sigma_{y}/2}\left|e\right\rangle
\end{array}\right\} \\
 & = & D\left(t\right),
\end{eqnarray}
where $\theta=\pi/2$ and $\sigma_{y}=i\left(\left|g\right\rangle \left\langle e\right|-\left|e\right\rangle \left\langle g\right|\right)$.
With the help of Eq.~(\ref{eq:ele_red_den}), we obtain the decoherence
factor as
\begin{equation}
D\left(t\right)\!\!=\!\!\mathrm{Re}\!\left\{ \!\exp\!\left[\!\left(\! W_{\sigma}^{*}\!\left(t\right)\! W_{\sigma'}\!\left(t\right)\!+\! J_{3,\sigma\sigma'}\!\left(t\right)\!-\!1\!\right)\!\left|\alpha\right|^{2}\!\right]\!\right\} \!.\!\!
\end{equation}

\end{document}